\documentclass[10pt,twocolumn]{IEEEtran}

\usepackage{amsmath,amssymb,epsfig,psfrag,cite,subfigure}
\include{macros}
\usepackage{graphicx}
\usepackage{epstopdf}

\usepackage{lipsum,multicol}

\usepackage{algorithm}
\usepackage{epsfig,psfrag}
\usepackage{color}
\usepackage{url}
\usepackage{mathtools,xparse}

\usepackage{gensymb}
\usepackage[multiple]{footmisc}

\usepackage{bbm}

\newcommand{\CRLB}{{\rm{CRLB}}}

\newcommand{\Tsi}{{T_{{\rm{s}},i}}}
\newcommand{\Tsik}{{T^k_{{\rm{s}},i}}}

\newcommand{\rmE}{{\rm{E}}}

\newcommand{\NL}{N_{\rm{L}}}
\newcommand{\NV}{N_{\rm{V}}}

\newcommand{\JF}{{\rm{\bf{J}}_{\rm{F}}}}

\newcommand{\Rj}{{\boldsymbol{R}}_j}
\newcommand{\RjT}{{\boldsymbol{R}}_j^T}
\newcommand{\EE}{{\boldsymbol{E}}}
\newcommand{\hj}{{\boldsymbol{h}}_j}
\newcommand{\hjT}{{\boldsymbol{h}}_j^T}
\newcommand{\hjh}{{\widehat{\boldsymbol{h}}}_j}

\newcommand{\bl}{{\boldsymbol{l}}}

\newcommand{\bvp}{{\boldsymbol{\varphi}}}

\newcommand{\bvphat}{{\hat{\bvp}}}

\newcommand{\mtC}{{\mathcal{C}}}

\newcommand{\expectation}{{\mathbb{E}}}

\DeclarePairedDelimiter{\norm}{\lVert}{\rVert}

\newcommand{\lr}{{\boldsymbol{l}_{\mathrm{r}}}}
\newcommand{\lrh}{{\widehat{\boldsymbol{l}}_{\mathrm{r}}}}
\newcommand{\lt}[1]{{\boldsymbol{l}^{#1}_{\mathrm{t}}}}

\newcommand{\nr}{{\boldsymbol{n}_{\mathrm{r}}}}
\newcommand{\nt}[1]{{\boldsymbol{n}^{#1}_{\mathrm{t}}}}

\newcommand{\lrs}[1]{{l_{\mathrm{r},#1}}}
\newcommand{\lts}[2]{{l^{#1}_{\mathrm{t},#2}}}

\newcommand{\nrs}[1]{{n_{\mathrm{r},#1}}}
\newcommand{\nts}[2]{{n^{#1}_{\mathrm{t},#2}}}

\newcommand{\Jsyni}{{\boldsymbol{\mathrm{J}}_{\mathrm{sce1}}^{-1}}}
\newcommand{\Jsyn}{{\boldsymbol{\mathrm{J}}_{\mathrm{sce1}}}}
\newcommand{\Jasyi}{{\boldsymbol{\mathrm{J}}_{\mathrm{sce2}}^{-1}}}
\newcommand{\Jasy}{{\boldsymbol{\mathrm{J}}_{\mathrm{sce2}}}}
\newcommand{\JsyniT}{{\boldsymbol{\mathrm{J}}_{\mathrm{sce3}}^{-1}}}
\newcommand{\JsynT}{{\boldsymbol{\mathrm{J}}_{\mathrm{sce3}}}}

\newcommand{\Jv}{{\boldsymbol{\mathrm{J}}(\bvp)}}
\newcommand{\Jvi}{{\boldsymbol{\mathrm{J}}^{-1}(\bvp)}}

\newcommand{\JA}{{\boldsymbol{\mathrm{J_A}}}}
\newcommand{\JB}{{\boldsymbol{\mathrm{J_B}}}}
\newcommand{\JBT}{{\boldsymbol{\mathrm{J}}}_{\boldsymbol{\mathrm{B}}}^{T}}

\newcommand{\JD}{{\boldsymbol{\mathrm{J_D}}}}
\newcommand{\JDi}{{\boldsymbol{\mathrm{J}}}_{\boldsymbol{\mathrm{D}}}^{-1}}

\newcommand{\JAt}{\tilde{{\boldsymbol{\mathrm{J}}}}_{\boldsymbol{\mathrm{A}}}}
\newcommand{\JBt}{\tilde{{\boldsymbol{\mathrm{J}}}}_{\boldsymbol{\mathrm{B}}}}

\newcommand{\JDt}{\tilde{{\boldsymbol{\mathrm{J}}}}_{\boldsymbol{\mathrm{D}}}}
\newcommand{\JDti}{\tilde{{\boldsymbol{\mathrm{J}}}}_{\boldsymbol{\mathrm{D}}}^{-1}}

\newcommand{\btau}{{\boldsymbol{\tau}}}
\newcommand{\btauh}{\widehat{{\boldsymbol{\tau}}}}

\newcommand{\hjk}{{\boldsymbol{h}}^{k}_{j}}
\newcommand{\hjkh}{\widehat{{\boldsymbol{h}}}^{k}_{j}}
\newcommand{\Rjk}{{\boldsymbol{R}}^{k}_{j}}
\newcommand{\Ek}{{\boldsymbol{E}}^{k}}

\begin{document}
%
\title{Distance and Position Estimation in Visible Light Systems with RGB LEDs}

\author{Ilker Demirel and\thanks{I. Demirel and S. Gezici are with the Department of Electrical and Electronics Engineering, Bilkent University, 06800, Ankara, Turkey, Tel: +90-312-290-3139, Fax: +90-312-266-4192, Emails: ilkerd@ee.bilkent.edu.tr, gezici@ee.bilkent.edu.tr} Sinan Gezici\thanks{Part of this research was presented at the IEEE 30th Annual International Symposium on Personal, Indoor and Mobile Radio Communications (PIMRC), Istanbul, Turkey, Sep. 8-11, 2019 \cite{IlkerPIMRC}.}}

\maketitle

\begin{abstract}
  In this manuscript, distance and position estimation problems are investigated for visible light positioning (VLP) systems with red-green-blue (RGB) light emitting diodes (LEDs). The accuracy limits on distance and position estimation are calculated in terms of the Cram\'{e}r-Rao lower bound (CRLB) for three different scenarios. Scenario~1 and Scenario~2 correspond to synchronous and asynchronous systems, respectively, with known channel attenuation formulas at the receiver. In Scenario~3, a synchronous system is considered but channel attenuation formulas are not known at the receiver. The derived CRLB expressions reveal the relations among distance/position estimation accuracies in the considered scenarios and lead to intuitive explanations for the benefits of using RGB LEDs.
  In addition, maximum likelihood (ML) estimators are derived in all scenarios, and it is shown that they can achieve close performance to the CRLBs in some cases for sufficiently high source optical powers.

\end{abstract}


%
\IEEEpeerreviewmaketitle

\section{Introduction}\label{sec:Intro}


Visible light positioning (VLP) systems have attracted significant attention in recent studies due to their low-cost and high-accuracy characteristics (\cite{SurveyVLPprocIEEE,MFK_ProcIEEE}, and references therein). In addition, they incur very low deployment cost as they are already employed for illumination. VLP systems with high localization accuracy can facilitate various applications such as real-time robot control, patient monitoring, and warehouse management \cite{VLP_Roadmap,SurveyVLProto,VLPind40}.

Among various theoretical and experimental studies related to VLP systems, a group of them focuses on determination of accuracy limits related to distance and position estimation \cite{CRB_TOA_VLC,VLP_CRLB_RSS,MFK_CRLB,IG_RSS_AOA_VLC,Erdal_CL_2015,ZZB_MFK,Direct_TCOM}. Accuracy limits provide theoretical performance bounds for a large class of estimators (such as unbiased estimators) and they can present guidelines for system design under specific accuracy requirements. In \cite{CRB_TOA_VLC}, the Cram\'{e}r-Rao lower bound (CRLB) is obtained for distance estimation based on the time-of-arrival (TOA) parameter in a synchronous VLP system and the dependence of the CRLB on various system parameters is investigated. In \cite{VLP_CRLB_RSS}, the CRLB is derived for distance estimation in an asynchronous VLP system, where the distance related information in the received signal strength (RSS) parameter is utilized. The work in \cite{MFK_CRLB} focuses on the distance estimation problem for both synchronous and asynchronous VLP systems, and considers the cases of known and unknown channel attenuation formulas at the visible light communication (VLC) receiver. It is shown that the distance related information contained in the TOA parameter (which can be utilized in the presence of synchronization) increases with the effective bandwidth of the transmitted optical waveform. Therefore, synchronous VLP systems can provide performance improvements over asynchronous ones only for sufficiently high effective bandwidths.


Regarding the position estimation problem in visible light systems, both theoretical limits and practical estimators are investigated in a multitude of studies such as \cite{Direct_TCOM,Guvenc_hybrid,VLP_RSS_DP,Steendam_Aperture_JLT_2017,DMT_VLC,VLP_DP_ICASSP,
zhang2014asynchronous,VLP_OFDM,MultiPD_VLC_2016}. In \cite{Guvenc_hybrid}, the CRLB is derived for three-dimensional localization in an indoor VLP system based on RSS information by considering a generic configuration for LED transmitters and VLC receiver. In \cite{VLP_RSS_DP} and \cite{Steendam_Aperture_JLT_2017}, two-dimensional RSS-based localization is studied under the assumption of a known receiver height, and an analytical CRLB expression is derived in the considered setting. The work in \cite{Direct_TCOM} provides the CRLBs for three dimensional position estimation in synchronous and asynchronous VLP systems by employing RSS and/or TOA parameters. In addition, it presents the maximum likelihood (ML) estimators for synchronous and asynchronous settings by employing direct and two-step positioning approaches. Instead of the RSS and TOA parameters, angle-of-arrival (AOA), time-difference-of-arrival (TDOA), or a combination of multiple parameters are employed in \cite{Guvenc_hybrid,GuvencWAMI15,TDOA_VLC,ThreeD_AOARSS,VLP_latency,Ertan_TDOA_VLP} for position estimation in visible light systems. For example, both AOA and RSS parameters are utilized in \cite{Guvenc_hybrid} to perform three-dimensional localization of VLC receivers. By taking a direct positioning approach, \cite{VLP_DP_ICASSP} proposes an asynchronous VLP system in which a Bayesian signal model is constructed for position estimation based on the entire received signal from multiple LEDs in the presence of obstruction of signals from several LEDs.

Although the theoretical limits on distance and position estimation and corresponding ML estimators are investigated for VLP systems with white LEDs in \cite{CRB_TOA_VLC,VLP_CRLB_RSS,MFK_CRLB,IG_RSS_AOA_VLC,Erdal_CL_2015,Direct_TCOM}, they are not available for VLP systems with red-green-blue (RGB) LEDs in the literature. Since RGB LEDs can provide additional benefits for visible light systems \cite{ColorShiftKeying_JLT14,RGBref1}, analysis of theoretical limits and derivation of ML estimators are crucial for VLP systems with RGB LEDs, as well. The aim of this manuscript is to provide a detailed analysis of the position estimation problem in visible light systems with RGB LEDs. We focus on three scenarios where Scenario~1 and Scenario~2 correspond to synchronous and asynchronous systems, respectively, with known channel attenuation formulas at the receiver. In Scenario~3, a synchronous system is considered but channel attenuation formulas are not known at the receiver. For all of these scenarios, we first focus on the distance estimation problem for VLP systems with RGB LEDs by considering a specific setting with a known VLC receiver height, and derive the CRLBs and ML estimators. Then, we consider the generic three-dimensional localization problem for visible light systems with RGB LEDs and derive the CRLBs and ML estimators. The provided CRLB expressions and ML estimators generalize the ones in the literature \cite{MFK_CRLB,Direct_TCOM} as there exist three parallel channels in RGB based VLP systems. In addition, Scenario~3, which is not considered in \cite{Direct_TCOM}, is investigated for VLP systems with both white and RGB LEDs. The main contributions and novelty of this manuscript can be summarized as follows:
\begin{itemize}
  \item The CRLBs and the ML estimators are derived for \emph{distance} estimation in VLP systems with RGB LEDs for the first time in the literature. The obtained results generalize those in \cite{MFK_CRLB} to VLP systems with RGB LEDs and reveal the benefits of employing RGB LEDs for distance estimation.
  \item The CRLBs and the ML estimators are derived for generic three-dimensional \emph{position} estimation in VLP systems with RGB LEDs for the first time in the literature. In this way, not only the results in \cite{Direct_TCOM} are extended to VLP systems with RGB LEDs but also a synchronous scenario with unknown channel attenuation formulas (Scenario~3) is investigated, which is not considered in \cite{Direct_TCOM}.
  \item Via the derived CRLB expressions, the relations among the distance/position estimation accuracies are revealed in the considered scenarios and the benefits of using RGB LEDs can be quantified.
\end{itemize}
In addition, numerical examples are provided to illustrate the theoretical results and to compare the performance of the ML estimators against the corresponding CRLBs in various scenarios. (In the conference version of this manuscript \cite{IlkerPIMRC}, only the \emph{distance} estimation problem was investigated for VLP systems with RGB LEDs.)

The remainder of the manuscript is organized as follows: Section~\ref{sec:Model} introduces the VLP system model with RGB LEDs. Then, the derivations of the CRLBs and ML estimators for distance estimation are performed in Section~\ref{sec:DistEst}. In Section~\ref{sec:PosEst}, the general case of three-dimensional localization is investigated by deriving CRLBs and ML estimators. The numerical examples are presented in Section~\ref{sec:Nume}, which are followed by the concluding remarks in Section~\ref{sec:Conc}.


\section{System Model}\label{sec:Model}

Consider a VLP system that consists of $\NL$ LED transmitters at known locations (e.g., on the ceiling of a room) and a VLC receiver at an unknown location. The VLC receiver estimates its location by utilizing the signals emitted by the LED transmitters (i.e., self-positioning  \cite{Sinan_Survey}). Let $\lr\in\mathbb{R}^3$ and $\lt{k}\in\mathbb{R}^3$ represent the locations of the VLC receiver and the $k$th LED transmitter, respectively, where $k\in\{1,\ldots,\NL\}$. Each LED transmitter can emit red, green, and blue signals (colors), which are denoted by $s^k_i(t)$ for $i\in\mtC$ and $k\in\{1,\ldots,\NL\}$ with
\begin{equation}\label{eq:Cset}
\mtC\triangleq\{r,g,b\}
\end{equation}
The VLC receiver processes the incoming optical signals from the LED transmitters via three parallel photodetectors (PDs) corresponding to red, green, and blue signals. It is assumed that a certain type of a multiple access protocol is employed at the VLC receiver so that signals from the LED transmitters can be processed separately \cite{Direct_TCOM,VLC_Survey}. Accordingly, the following electrical signals are observed at the VLC receiver:
\begin{gather}\label{eq:RecSig}
y^k_j(t)=\sum_{i\in\mtC}h^k_{j,i}\,s^k_i\big(t-\tau^k\big)+\eta^k_j(t)
\end{gather}
for $k\in\{1,\ldots,\NL\}$, $j\in\mtC$ and $t\in[T^k_{1,j},T^k_{2,j}]$, where $T^k_{1,j}$ and $T^k_{2,j}$ specify the observation interval for PD~$j$ related to the signal coming from the $k$th LED transmitter, $h^k_{j,i}$ is the overall channel attenuation factor for PD~$j$ and the $i$th signal (color) of the $k$th LED transmitter ($h^k_{j,i}>0$),
$\tau^k$ is the TOA parameter related to the $k$th LED transmitter, and $\eta^k_j(t)$ is the noise at PD~$j$ during the reception of the signal from the $k$th LED transmitter.


The noise terms $\eta^k_j(t)$ in \eqref{eq:RecSig} are modeled as zero-mean white Gaussian random processes with a spectral density level of $\sigma_j^2$, which are assumed to be independent for all $k\in\{1,\ldots,\NL\}$ (due to the use of a multiple access protocol \cite{Direct_TCOM,VLC_Survey}) and for all $j\in\mtC$ (due to the processing at different branches of the VLC receiver). The transmitted signals $s^k_i(t)$ are nonzero over an interval of $[0,\Tsik]$ for $k\in\{1,\ldots,\NL\}$ and $i\in\mtC$, and they are assumed to be known by the VLC receiver.
Also, the TOA parameter in \eqref{eq:RecSig} can be expressed as
\begin{gather}\label{eq:tau}
\tau^k=\frac{\|\lr-\lt{k}\|_2}{c}+\Delta^k
\end{gather}
where $c$ denotes the speed of light and $\Delta^k$ is the time offset between the clocks of the $k$th LED transmitter and the VLC receiver. For synchronous VLP systems, $\Delta^k=0$ for all $k\in\{1,\ldots,\NL\}$, whereas $\Delta^k$'s are unknown parameters for asynchronous VLP systems. As in \cite{MFK_CRLB}, it is assumed that a coarse acquisition is performed such that the signal component in \eqref{eq:RecSig} resides completely in the observation interval $[T^k_{1,j},T^k_{2,j}]$ for $k\in\{1,\ldots,\NL\}$ and $j\in\mtC$.

As in \cite{CRB_TOA_VLC,VLP_CRLB_RSS,Direct_TCOM}, a line-of-sight scenario is considered and the overall channel attenuation factors in \eqref{eq:RecSig} are modeled as \cite{WirelessInfComm_97,zhang2014asynchronous,Chvojka15,EPSILON}
\begin{gather} \label{eq:hji}
h^k_{j,i} = - \frac{(m^k + 1)A_j}{2 \pi} \frac{[(\lr - \lt{k})^T \nt{k}]^{m^k} (\lr - \lt{k})^T \nr}{\big\|\lr - \lt{k}\big\|^{m^k + 3}}\,\tilde{R}_{j,i}
\end{gather}
for $i,j\in\mtC$ and $k\in\{1,\ldots,\NL\}$, where $m^k$ is the Lambertian order for the $k$th LED transmitter, $A_j$ is the area of PD~$j$, $\nr$ and $\nt{k}$ denote the orientation vectors for the VLC receiver and the $k$th LED transmitter, respectively, and $\tilde{R}_{j,i}$ is the responsivity of PD $j$ to the $i$th signal (color).


\section{Distance Estimation}\label{sec:DistEst}

Before investigating the general case of three-dimensional localization, we first focus on a special scenario in which the VLC receiver performs distance (range) estimation with each of the LED transmitters, and then determines its two-dimensional location based on those distance estimates \cite{MFK_CRLB}. In this scenario, accuracy of distance estimation is the main factor that determines the accuracy of location estimation \cite[Sec.~VI]{MFK_CRLB}. Therefore, the purpose in this section is to determine the accuracy limits of distance estimation for VLP systems with RGB LEDs, which has not been investigated in the literature.


\begin{figure*}
\centering
\includegraphics[width=5.5in]{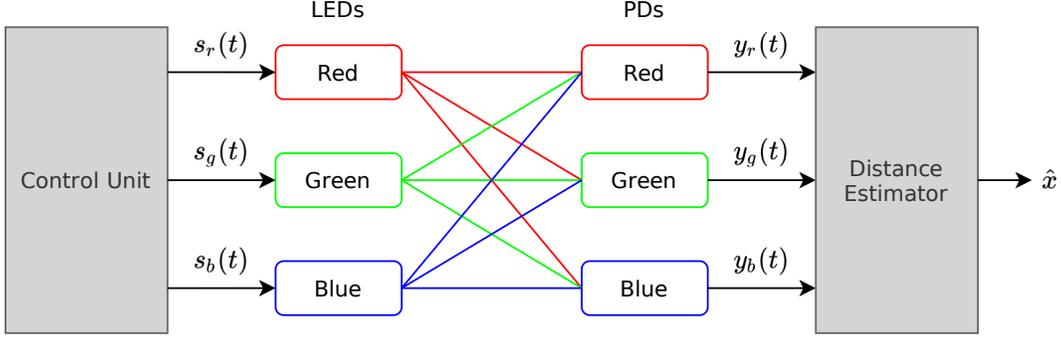}
\caption{For distance estimation, VLP system model with RGB LEDs.}
\label{fig:system}
\end{figure*}

As the aim is to estimate the distance between each LED transmitter and the VLC receiver, we focus on one LED transmitter and drop the index $k$ (superscript $k$) from the parameters in Section~\ref{sec:Model}. The system model for this scenario is shown in Fig.~\ref{fig:system}, and the distance between the LED transmitter and the VLC receiver is represented by $x$, which is given by $x=\|\lr-\bl_t\|_2$. As in \cite{MFK_CRLB,VLP_CRLB_RSS,CRB_TOA_VLC,LED_MultiRec}, it is assumed in this section that the LED transmitter points downwards (which is a common scenario) and the VLC receiver points upwards such that $\nr=-\nt{}=[0~0~1]$ and $-(\lr - \lt{})^T \nr=(\lr - \lt{})^T \nt{}=\tilde{h}$ in \eqref{eq:hji},
where $\tilde{h}$ denotes the height of the LED transmitter relative to the VLC receiver. In addition, it is assumed similarly to \cite{MFK_CRLB,VLP_CRLB_RSS,CRB_TOA_VLC,LED_MultiRec} that $\tilde{h}$ is known by the VLC receiver; that is, possible locations of the VLC receiver are confined to a two-dimensional plane (e.g., to the floor of a room). This assumption is made (only in this section) to provide intuitive and clear explanations for accuracy limits of distance estimation in VLP systems with RGB LEDs. (It also holds in many practical scenarios; e.g., when the VLC receiver is attached to a warehouse vehicle or a robot; see Fig.~3 in \cite{VLP_Roadmap}). Under these assumptions, the channel attenuation factors in \eqref{eq:hji} can be expressed as
\begin{gather}\label{eq:hji_2}
h_{j,i}=\frac{A_j(m+1)}{2\pi x^2}\left(\frac{\tilde{h}}{x}\right)^{m+1}\tilde{R}_{j,i}
\triangleq \gamma_{j,i}\,x^{-m-3}
\end{gather}
for $i,j\in\mtC$, where $\gamma_{j,i}\triangleq {A_j(m+1)\tilde{h}^{m+1}\tilde{R}_{j,i}}/{(2\pi)}$.

\subsection{Theoretical Limits for Distance Estimation}\label{sec:Limits}

In this section, accuracy limits of distance estimation are investigated for VLP systems with RGB LEDs under various scenarios.

\subsubsection{Scenario~1: Synchronous System with Known Channel Attenuation Formula}\label{sec:Case1}

In this scenario, it is assumed that the LED transmitter and the VLC receiver are synchronized; hence, $\Delta=0$ in \eqref{eq:tau}. In addition, the VLC receiver is assumed to know the channel attenuation formula in \eqref{eq:hji_2} with $\gamma_{j,i}$'s being known constants. (In practice, $\gamma_{j,i}$'s can be learned via calibration by placing the VLC receiver at known distances.) Under these assumptions, the log-likelihood function corresponding to the received signals in \eqref{eq:RecSig} (for only one LED transmitter) can be expressed, based on \eqref{eq:tau} with $\Delta=0$ and \eqref{eq:hji_2}, as follows:
\begin{align}\label{eq:loglike1}
\Lambda(x) = \tilde{K} - &\sum_{j\in\mtC}  \frac{1}{2\sigma_j^2}\int_{T_{1,j}}^{T_{2,j}}
\bigg(y_j(t)
\\\nonumber
&-\sum_{i\in\mtC}\gamma_{j,i}\,x^{-m-3}\,s_i\Big(t-\frac{x}{c}\Big)\bigg)^2dt
\end{align}
where $\tilde{K}$ is a constant that does not depend on $x$.

The CRLB provides a lower limit on MSEs of unbiased estimators and can be obtained from the log-likelihood function as follows:
\begin{gather}\label{eq:CRLB1_1}
\CRLB_1=\left(\rmE\left\{\left(\frac{d\Lambda(x)}{dx}\right)^2\right\}\right)^{-1}
\end{gather}
From \eqref{eq:loglike1}, the CRLB in \eqref{eq:CRLB1_1} can be derived as
\begin{align}\nonumber
&\CRLB_1=\Bigg(\sum_{j\in\mtC}\frac{1}{\sigma_j^2}\int_{T_{1,j}}^{T_{2,j}}
\bigg[x^{-m-4}\sum_{i\in\mtC}\gamma_{j,i}
\\\label{eq:CRLB1_2}
&\times\Big((m+3)s_i(t-x/c)+(x/c)s'_i(t-x/c)\Big)\bigg]^2dt
\Bigg)^{-1}
\end{align}
where $s'_i(t)$ denotes the derivative of $s_i(t)$. After some manipulation, \eqref{eq:CRLB1_2} can be expressed as
\begin{align}\nonumber
\CRLB_1=&\bigg((m+3)^2x^{-2m-8}\kappa+\frac{x^{-2m-6}}{c^2}\kappa''
\\\label{eq:CRLB1_3}
&+\frac{2(m+3)}{c}x^{-2m-7}\kappa'\bigg)^{-1}
\end{align}
where
\begin{align}\label{eq:kappa}
\kappa&\triangleq\sum_{j\in\mtC}\frac{1}{\sigma_j^2}\sum_{i\in\mtC}\sum_{l\in\mtC}\gamma_{j,i}\,\gamma_{j,l}\,E_{i,l}\\\label{eq:kappa1}
\kappa'&\triangleq\sum_{j\in\mtC}\frac{1}{\sigma_j^2}\sum_{i\in\mtC}\sum_{l\in\mtC}\gamma_{j,i}\,\gamma_{j,l}\,E'_{i,l}\\\label{eq:kappa2}
\kappa''&\triangleq\sum_{j\in\mtC}\frac{1}{\sigma_j^2}\sum_{i\in\mtC}\sum_{l\in\mtC}\gamma_{j,i}\,\gamma_{j,l}\,E''_{i,l}
\end{align}
with
\begin{align}\label{eq:Eil}
E_{i,l}&\triangleq\int_{-\infty}^{\infty}s_i(t)s_l(t)dt\\\label{eq:Eil1}
E'_{i,l}&\triangleq\int_{-\infty}^{\infty}s_i(t)s'_l(t)dt\\\label{eq:Eil2}
E''_{i,l}&\triangleq\int_{-\infty}^{\infty}s'_i(t)s'_l(t)dt
\end{align}

Various observations can be made based on \eqref{eq:CRLB1_3}--\eqref{eq:Eil2}. First, as expected, there is contribution to accuracy from all three colors. Second, distance related information contained in both the received signal strength (RSS) parameter and the TOA parameter is utilized in this scenario. To see this more clearly, suppose that the same intensity levels are transmitted from all the LEDs such that $s_r(t)=s_g(t)=s_b(t)$. Also, assume that $s_i(0)=s_i(\Tsi)$ for $i\in\mtC$, which is commonly the case for practical signals. Then, it is obtained from \eqref{eq:kappa1} and \eqref{eq:Eil1} that $\kappa'=0$; hence, the CRLB in \eqref{eq:CRLB1_3} becomes $\CRLB_1=\big((m+3)^2x^{-2m-8}\kappa+{x^{-2m-6}}{c^{-2}}\kappa''\big)^{-1}$. In this expression, the first term comes from the information obtained from the RSS parameter based on the known channel attenuation formula (as $\kappa$ is related to the total received power) and the second term is due to the TOA parameter (since $\kappa''$ is related to the time resolution; equivalently, the effective bandwidth of the signals). As a final observation, it can be shown that the CRLB formula in \eqref{eq:CRLB1_3}--\eqref{eq:Eil2} covers the one in \cite{MFK_CRLB} as a special case if there exists only one LED at the transmitter and one PD at the receiver (cf.~\cite[Sec.~III-A]{MFK_CRLB}).

\subsubsection{Scenario~2: Asynchronous System with Known Channel Attenuation Formula}\label{sec:Case2}

In this scenario, it is assumed that the LED transmitter and the VLC receiver are asynchronous; namely, $\Delta$ in \eqref{eq:tau} is modeled as a deterministic unknown parameter. However, the VLC receiver is assumed to know the channel attenuation formula in \eqref{eq:hji_2} with $\gamma_{j,i}$'s being known constants. In this case, the log-likelihood function corresponding to the received signals in \eqref{eq:RecSig} can be expressed via \eqref{eq:hji_2} as
\begin{align}\label{eq:loglike2}
\Lambda(x,\tau) = \tilde{K} - &\sum_{j\in\mtC} \frac{1}{2\sigma_j^2}\int_{T_{1,j}}^{T_{2,j}}
\bigg(y_j(t)
\\\nonumber
&-\sum_{i\in\mtC}\gamma_{j,i}\,x^{-m-3}\,s_i(t-\tau)\bigg)^2dt
\end{align}
where $\tilde{K}$ is a constant that does not depend on $x$ or $\tau$. Then, the CRLB on distance estimation is given by\footnote{In \eqref{eq:CRLB2_1} and \eqref{eq:CRLB3_1}, $[{\bf{X}}]_{11}$ denotes the element of matrix ${\bf{X}}$ at row $1$ and column $1$.}
\begin{gather}\label{eq:CRLB2_1}
\CRLB_2=\left[\JF^{-1}\right]_{11}
\end{gather}
where $\JF$ is the Fisher information matrix (FIM) defined as
\begin{gather}\label{eq:FIM2}
\JF=\begin{bmatrix}
\rmE\left\{\left(\frac{\partial\Lambda(x,\tau)}{\partial x}\right)^2\right\}&\rmE\left\{\frac{\partial\Lambda(x,\tau)}{\partial x}\frac{\partial\Lambda(x,\tau)}{\partial\tau}\right\}\\
\rmE\left\{\frac{\partial\Lambda(x,\tau)}{\partial x}\frac{\partial\Lambda(x,\tau)}{\partial\tau}\right\}&\rmE\left\{\left(\frac{\partial\Lambda(x,\tau)}{\partial\tau}\right)^2\right\}
\end{bmatrix}.
\end{gather}
From \eqref{eq:loglike2}, the elements of the FIM in \eqref{eq:FIM2} can be calculated after some manipulation as
\begin{gather}\label{eq:FIM2_2}
\JF=\begin{bmatrix}
(m+3)^2x^{-2m-8}\kappa & (m+3)x^{-2m-7}\kappa'
\\(m+3)x^{-2m-7}\kappa' & x^{-2m-6}\kappa''
\end{bmatrix}
\end{gather}
where $\kappa$, $\kappa'$, and $\kappa''$ are as defined in \eqref{eq:kappa}--\eqref{eq:kappa2}.

Based on \eqref{eq:CRLB2_1} and \eqref{eq:FIM2_2}, the CRLB on distance estimation can be calculated as
\begin{gather}\label{eq:CRLB2_2}
\CRLB_2=\frac{\kappa''x^{2m+8}}{(m+3)^2(\kappa\,\kappa''-(\kappa')^2)}\,\cdot
\end{gather}

By comparing \eqref{eq:CRLB1_3} and \eqref{eq:CRLB2_2}, it is noted that only the RSS parameter is utilized in this scenario since there is no synchronization between the transmitter and the receiver. In particular, if $\kappa'=0$, then the CRLB in \eqref{eq:CRLB2_2} becomes $\CRLB_2=((m+3)^2x^{-2m-8}\kappa)^{-1}$, which corresponds to the first term in \eqref{eq:CRLB1_3}, as expected. Also, it is noted that the CRLB formula in \eqref{eq:CRLB2_2} covers the one in \cite[Sec.~III-B]{MFK_CRLB} as a special case if there exists only one LED at the transmitter and one PD at the receiver.

\subsubsection{Scenario~3: Synchronous System with Unknown Channel Attenuation Formula}\label{sec:Case3}

In the final scenario, it is assumed that the LED transmitter and the VLC receiver are synchronized (i.e., $\Delta=0$ in \eqref{eq:tau}) but the VLC receiver does not know the channel attenuation formula in \eqref{eq:hji_2}. Then, the log-likelihood function corresponding to the received signals in \eqref{eq:RecSig} can be expressed, based on \eqref{eq:tau} with $\Delta=0$, as follows:
\begin{align}\label{eq:loglike3}
\Lambda(\bvp) = \tilde{K} - &\sum_{j\in\mtC}  \frac{1}{2\sigma_j^2}\int_{T_{1,j}}^{T_{2,j}}
\bigg(y_j(t)
\\\nonumber
&-\sum_{i\in\mtC}h_{j,i}\,s_i\Big(t-\frac{x}{c}\Big)\bigg)^2dt
\end{align}
where $\bvp=[x~h_{r,r}~h_{r,g}~h_{r,b}~h_{g,r}~h_{g,g}~h_{g,b}~h_{b,r}~h_{b,g}~h_{b,b}]^T$ is the vector of unknown parameters and $k$ is a constant that does not depend on $\bvp$.

In this scenario, the CRLB on distance estimation is stated as
\begin{gather}\label{eq:CRLB3_1}
\CRLB_3=\left[\JF^{-1}\right]_{11}
\end{gather}
where $\JF$ is the FIM, which has a size of $10\times10$. The elements of $\JF$ are specified as follows:
\begin{gather}\label{eq:FIM3}
\JF=\begin{bmatrix}
{\textrm{A}}&{\bf{B}}\\
{\bf{B}}^T&{\bf{D}}
\end{bmatrix}
\end{gather}
where ${\textrm{A}}=\rmE\left\{\left(\frac{\partial\Lambda(\bvp)}{\partial x}\right)^2\right\}$, ${\bf{B}}$ is a $1\times9$ vector given by ${\bf{B}}=\left[\rmE\left\{\frac{\partial\Lambda(\bvp)}{\partial x}\frac{\partial\Lambda(\bvp)}{\partial h_{l,k}}\right\}\right]$ for $l,k\in\mtC$, and ${\bf{D}}$ is a $9\times9$ matrix defined as ${\bf{D}}=\left[\rmE\left\{\frac{\partial\Lambda(\bvp)}{\partial h_{l,k}}\frac{\partial\Lambda(\bvp)}{\partial h_{n,m}}\right\}\right]$ for $l,k,n,m\in\mtC$. Based on \eqref{eq:loglike3}, the elements of $\JF$ in \eqref{eq:FIM3} can be specified as follows:
\begin{align}\label{eq:A}
{\textrm{A}}&=\frac{1}{c^2}\sum_{j\in\mtC}\frac{1}{\sigma_j^2}
\sum_{i\in\mtC}\sum_{l\in\mtC}h_{j,i}\,h_{j,l}\,E''_{i,l}\\\label{eq:B}
{\bf{B}}&=\left[\frac{1}{c\,\sigma_l^2}\sum_{i\in\mtC}\,h_{l,i}\,E'_{k,i}\right],~~l,k\in\mtC\\\label{eq:C}
{\bf{D}}&=\left[\frac{E_{k,m}\mathbbm{1}_{\{l=n\}}}{\sigma_l^2}\right],~~l,k,n,m\in\mtC
\end{align}
where $E_{k,m}$, $E'_{k,i}$, and $E''_{i,l}$ are as defined in \eqref{eq:Eil}--\eqref{eq:Eil2}, and $\mathbbm{1}_{\{l=n\}}$ is the indicator function, which is equal to one if $l=n$ and zero otherwise.

From \eqref{eq:FIM3}--\eqref{eq:C}, the CRLB in \eqref{eq:CRLB3_1} can be obtained as
\begin{gather}\label{eq:CRLB3_2}
\CRLB_3=\left({\textrm{A}}-{\bf{B}}{\bf{D}}^{-1}{\bf{B}}^T\right)^{-1}\,.
\end{gather}
It is noted from \eqref{eq:C} that the $9\times9$ matrix ${\bf{D}}$ has a block diagonal structure; hence, the calculation of \eqref{eq:CRLB3_2} requires inversion of three $3\times3$ matrix blocks.

In this scenario, the distance related information in the TOA parameter is utilized since the system is synchronous but the channel attenuation formula is unknown. As a special case, if $E'_{k,i}=0$ for all $k,i\in\mtC$, then ${\bf{B}}=\bf{0}$ and $\CRLB_3=1/{\textrm{A}}$. In this case, unknown channel attenuation factors, $h_{j,i}$'s, do not affect the distance estimation accuracy. In all other cases, distance estimation accuracy is affected by the presence of unknown channel attenuation factors (as they influence how accurately the TOA information can be extracted).

\subsection{ML Estimators for Distance Estimation}\label{sec:ML}

In this section, the ML estimators are derived for the scenarios considered in the previous subsection.

\subsubsection{ML Estimator for Scenario~1}\label{sec:ML1}

The ML estimator in Scenario~1 is stated as
\begin{gather}\label{eq:ML1}
\widehat{x}_1=\underset{x}{\arg\max}~\Lambda(x)
\end{gather}
where $\Lambda(x)$ is given by \eqref{eq:loglike1}. Based on the expression in \eqref{eq:loglike1}, the ML estimator in \eqref{eq:ML1} can be specified, after some manipulation, as
\begin{align}\nonumber
\widehat{x}_1=\underset{x}{\arg\max}~&x^{-m-3}
\sum_{j\in\mtC}\frac{1}{\sigma_j^2}
\sum_{i\in\mtC}
\gamma_{j,i}R_{y_j,s_i}\bigg(\frac{x}{c}\bigg)
\\\label{eq:ML1_2}
&-0.5\,x^{-2m-6}\kappa
\end{align}
where $\kappa$ is as in \eqref{eq:kappa} and
\begin{gather}\label{eq:Rys}
R_{y_j,s_i}(\tau)\triangleq \int_{T_{1,j}}^{T_{2,j}}y_j(t)s_i(t-\tau)dt\,.
\end{gather}
As noted from \eqref{eq:ML1_2}, a one-dimensional search is required to obtain the distance estimate. However, for each possible distance, correlations of the received signals are performed with delayed versions of the transmitted signals from the LEDs (see \eqref{eq:Rys}).

\subsubsection{ML Estimator for Scenario~2}\label{sec:ML2}

The ML estimator in Scenario~2 is defined as
\begin{gather}\label{eq:ML2}
(\widehat{x}_2,\widehat{\tau})=\underset{(x,\tau)}{\arg\max}~\Lambda(x,\tau)
\end{gather}
where $\Lambda(x,\tau)$ is as in \eqref{eq:loglike2}. After some manipulation, the ML estimator in \eqref{eq:ML2} can be expressed as
\begin{align}\nonumber
(\widehat{x}_2,\widehat{\tau})=\underset{(x,\tau)}{\arg\max}~&x^{-m-3}
\sum_{j\in\mtC}\frac{1}{\sigma_j^2}
\sum_{i\in\mtC}
\gamma_{j,i}R_{y_j,s_i}(\tau)
\\\label{eq:ML2_2}
&-0.5\,x^{-2m-6}\kappa
\end{align}
where $R_{y_j,s_i}(\tau)$ is given by \eqref{eq:Rys}. From \eqref{eq:ML2_2}, $\widehat{\tau}$ can be obtained as
\begin{gather}\label{eq:tauHat}
\widehat{\tau}=\underset{\tau}{\arg\max}
\sum_{j\in\mtC}\frac{1}{\sigma_j^2}
\sum_{i\in\mtC}
\gamma_{j,i}R_{y_j,s_i}(\tau)\,.
\end{gather}
Then, $\widehat{\tau}$ can be inserted into the objective function in \eqref{eq:ML2_2}, and derivatives with respect to $x$ can be calculated to show that $\widehat{x}_2$ is given by the following formula:
\begin{gather}\label{eq:x2Hat}
\widehat{x}_2=\left(\frac{1}{\kappa}\sum_{j\in\mtC}\frac{1}{\sigma_j^2}
\sum_{i\in\mtC}
\gamma_{j,i}R_{y_j,s_i}\big(\widehat{\tau}\big)\right)^{\frac{-1}{m+3}}
\end{gather}
where $\widehat{\tau}$ is as defined in \eqref{eq:tauHat} and $\sum_{j\in\mtC}\sigma_j^{-2}
\sum_{i\in\mtC}
\gamma_{j,i}R_{y_j,s_i}\big(\widehat{\tau}\big)$ is assumed to be positive.

From \eqref{eq:tauHat} and \eqref{eq:x2Hat}, it is noted that the ML distance estimation is performed in two steps in Scenario~2. In the first step, the TOA parameter is estimated. In the second step, this TOA estimate is used to determine the RSS level, which is then employed in the distance estimation process by utilizing the known channel attenuation formulas.

\subsubsection{ML Estimator for Scenario~3}\label{sec:ML3}

The ML estimator in Scenario~3 is defined as
\begin{gather}\label{eq:ML3}
\widehat{\bvp}=\underset{\bvp}{\arg\max}~\Lambda(\bvp)
\end{gather}
where $\Lambda(\bvp)$ is given by \eqref{eq:loglike3}. After some manipulation, the ML estimator in \eqref{eq:ML3} can be stated as
\begin{gather}\label{eq:ML3_2}
\widehat{\bvp}=\underset{\bvp}{\arg\max}~
\sum_{j\in\mtC}\frac{1}{\sigma_j^2}
\left(\hjT\Rj(x)-0.5\,\hjT\EE\,\hj\right)
\end{gather}
where $\hj\triangleq[h_{j,r}~h_{j,g}~h_{j,b}]^T$ and $\Rj(x)\triangleq[R_{y_j,s_r}(x)~R_{y_j,s_g}(x)~R_{y_j,s_b}(x)]^T$ with $R_{y_j,s_i}(x)$ being given by \eqref{eq:Rys}. In addition, $\EE$ in \eqref{eq:ML3_2} is a $3\times 3$ symmetric matrix described as
\begin{gather}\label{eq:Ematrix}
\EE=\begin{bmatrix}
E_{r,r}&E_{r,g}&E_{r,b}\\
E_{g,r}&E_{g,g}&E_{g,b}\\
E_{b,r}&E_{b,g}&E_{b,b}
\end{bmatrix}
\end{gather}
whose elements are as defined in \eqref{eq:Eil}. For simplicity of the derivations, $\EE$ is assumed to be positive definite in the remainder of the manuscript.

The gradient of the objective function in \eqref{eq:ML3_2} with respect to $\hj$ is calculated as $\sigma_j^{-2}(\Rj(x)-\EE\,\hj)$ for $j\in\mtC$. Then, the ML estimates for $\hj$ are obtained as $\hjh=\EE^{-1}\Rj(x)$ for $j\in\mtC$. Inserting these estimates into \eqref{eq:ML3_2}, the ML distance estimate is derived as
\begin{gather}\label{eq:ML3_3}
\widehat{x}_3=\underset{x}{\arg\max}\sum_{j\in\mtC}
\frac{1}{\sigma_j^2}\,\RjT(x)\EE^{-1}\Rj(x)\,.
\end{gather}
Similar to Scenario~1, a one-dimensional search is performed to obtain the distance estimate and for each possible distance, correlations of the received signals are calculated with delayed versions of the transmitted signals from the LEDs.

\subsubsection{Modified ML Estimator for Scenario~1}\label{sec:ML4}

To utilize the distance related information in the TOA parameter effectively, it can be required to sample the correlation function in \eqref{eq:Rys} at high rates. Otherwise, the performance of the ML estimator in \eqref{eq:ML1_2} may not get very close to the CRLB. To mitigate this problem, a modified version of the ML estimator can be designed as proposed in \cite{MFK_CRLB}. In particular, the ML estimate calculated from \eqref{eq:ML1_2} can be used as an input to the relation in \eqref{eq:x2Hat}; that is, the modified ML estimator can be obtained as
\begin{gather}\label{eq:x4Hat}
\widehat{x}_4=\left(\frac{1}{\kappa}\sum_{j\in\mtC}\frac{1}{\sigma_j^2}
\sum_{i\in\mtC}
\gamma_{j,i}R_{y_j,s_i}\bigg(\frac{\widehat{x}_1}{c}\bigg)\right)^{\frac{-1}{m+3}}
\end{gather}
where $\widehat{x}_1$ is the ML estimate in \eqref{eq:ML1_2}. This estimator is robust against sampling rate limitations, as observed in Section~\ref{sec:Nume}.


\section{Position Estimation}\label{sec:PosEst}

In this section, we consider a generic three-dimensional localization scenario in which the LED transmitters and the VLC receiver can have any orientations and locations. In particular, the aim is to estimate the location $\lr$ of the VLC receiver based on the received signals in \eqref{eq:RecSig}; namely, $y^k_j(t)$ for $t\in[T^k_{1,j},T^k_{2,j}]$, $j\in\mtC$, and $k\in\{1,\ldots,\NL\}$. Based on these received signals, the log-likelihood function can be stated as
\begin{align}\nonumber
\Lambda (\bvp) &= \tilde{K} -
\sum_{k=1}^{\NL}\sum_{j\in\mtC}\frac{1}{2\sigma_j^2}
\\\label{eqn:loglike}
&\times\int_{T^k_{1,j}}^{T^k_{2,j}}\bigg(y^k_j(t)-\sum_{i\in\mtC}h^k_{j,i}\,s^k_i\big(t-\tau^k\big)\bigg)^2dt
\end{align}
where $\bvp$ denotes the set of unknown parameters and $\tilde{K}$ is a constant that does not depend on the unknown parameters. The set of unknown parameters varies according to the considered scenario as specified below. The CRLB on the covariance matrix of any unbiased estimator $\hat{\bvp}$ of $\bvp$ can be expressed as
\cite{Poor}
\begin{gather}\label{crlb_expression}
\expectation \big\{ ( \bvphat - \bvp  ) ( \bvphat - \bvp  )^T \big\} \succeq \Jv^{-1}
\end{gather}
where $\Jv$ denotes the FIM for $\bvp$ and $\tilde{\boldsymbol{A}} \succeq \tilde{\boldsymbol{B}}$ means that $\tilde{\boldsymbol{A}} - \tilde{\boldsymbol{B}}$ is positive semidefinite. The FIM is computed as
\begin{gather}\label{eq:fim}
\Jv = \expectation \left\{ \left(  \nabla_{\bvp} \Lambda(\bvp) \right) \left( \nabla_{\bvp} \Lambda(\bvp) \right)^T \right\}
\end{gather}
where $\nabla_{\bvp}$ represents the gradient operator with respect to $\bvp$ and $\Lambda(\bvp)$ is the log-likelihood function in \eqref{eqn:loglike}.

\subsection{Theoretical Limits for Position Estimation}\label{sec:LimitsPos}

In this section, accuracy limits of position estimation are investigated for VLP systems with RGB LEDs under three different scenarios.

\subsubsection{Scenario~1: Synchronous System with Known Channel Attenuation Formula}\label{sec:Case1pos}

In this scenario, the clocks of the VLC receiver and the LED transmitters are synchronized (that is, $\Delta^k=0$ for all $k\in\{1,\ldots,\NL\}$ in \eqref{eq:tau}), and the VLC receiver knows the channel attenuation formula in \eqref{eq:hji}. Then, the set of unknown parameters in \eqref{eqn:loglike} becomes $\bvp=\lr=[\lrs{1}~\lrs{2}~\lrs{3}]^T$; i.e., the only unknown parameter is the location of the VLC receiver. For this scenario, the CRLB is given by the following proposition.

\textit{\textbf{Proposition~1}: In Scenario~1, the CRLB on the MSE of any unbiased estimator $\lrh$ for the location of the VLC receiver is given by}
\begin{gather}\label{eq:CRLBsync}
\expectation \big\{ \big\|\lrh - \lr \big\|^2 \big\}\geq\rm{trace}\big\{\Jsyni\big\}
\end{gather}
\textit{where}
\begin{align}\nonumber
[\Jsyn]_{n_1,n_2}&=\sum_{j\in\mtC}\frac{1}{\sigma_j^2}\sum_{k=1}^{\NL}\sum_{i\in\mtC}
\sum_{l\in\mtC}
\bigg(
\frac{\partial h^k_{j,i}}{\partial \lrs{n_1}}\frac{\partial h^k_{j,l}}{\partial \lrs{n_2}}E^k_{i,l}\\\nonumber
&\quad-\frac{\partial h^k_{j,i}}{\partial \lrs{n_1}}\frac{\partial \tau^k}{\partial \lrs{n_2}}h^k_{j,l}E'^{,k}_{i,l}\\\nonumber
&\quad-\frac{\partial \tau^k}{\partial \lrs{n_1}}\frac{\partial h^k_{j,l}}{\partial \lrs{n_2}}h^k_{j,i}E'^{,k}_{l,i}\\
&\quad+\frac{\partial \tau^k}{\partial \lrs{n_1}}\frac{\partial \tau^k}{\partial \lrs{n_2}}h^k_{j,l}h^k_{j,i}E''^{,k}_{i,l}
\bigg)\label{eq:Jsyc1}
\end{align}
\textit{for $n_1,n_2\in\{1,2,3\}$ with}
\begin{align}\label{eq:Eilpos}
E^k_{i,l}&\triangleq\int_{-\infty}^{\infty}s^k_i(t)s^k_l(t)dt\\\label{eq:Eil1pos}
E'^{,k}_{i,l}&\triangleq\int_{-\infty}^{\infty}s^k_i(t)(s^k_l(t))'dt\\\label{eq:Eil2pos}
E''^{,k}_{i,l}&\triangleq\int_{-\infty}^{\infty}(s^k_i(t))'(s^k_l(t))'dt
\\\label{eq:tauDer}
\frac{\partial \tau^k}{\partial \lrs{n}}&=\frac{\lrs{n}-\lts{k}{n}}{c\norm{\lr - \lt{k}}}
\\\label{eq:alpDer}
\frac{\partial h^k_{j,i}}{\partial \lrs{n}}&=-\frac{(m^k+1)A_j\tilde{R}_{j,i}}{2\pi}\bigg(
\frac{\big((\lr - \lt{k})^T\nt{k}\big)^{m^k-1}}{\norm{\lr - \lt{k}}^{m^k+3}}
\\\nonumber
&\hspace{-0.3cm}\times\big(m^k\,\nts{k}{n}(\lr - \lt{k})^T\nr
+\nrs{n}(\lr - \lt{k})^T\nt{k}\,\big)
\\\nonumber
&\hspace{-0.3cm}-\frac{(m^k+3)(\lrs{n}-\lts{k}{n})}{\norm{\lr - \lt{k}}^{m^k+5}}\big((\lr - \lt{k})^T\nt{k}\big)^{m^k}(\lr - \lt{k})^T\nr\bigg)
\end{align}

\indent\indent\textit{Proof:} In Scenario~1, the log-likelihood function in \eqref{eqn:loglike} is considered for $\bvp=\lr$, where $\tau^k$ is given by \eqref{eq:tau} with $\Delta^k=0$ for all $k\in\{1,\ldots,\NL\}$ and $h^k_{j,i}$ is as in \eqref{eq:hji}. Then, the elements of the FIM in \eqref{eq:fim} is stated as
\begin{gather}\label{eq:genericFIMsyc}
	[\Jv]_{n_1,n_2}=\expectation \left\{ \frac{\partial\Lambda(\bvp)}{\partial\lrs{n_1}}
	\frac{\partial\Lambda(\bvp)}{\partial\lrs{n_2}} \right\}
\end{gather}
for $n_1,n_2\in\{1,2,3\}$, where ${\partial\Lambda(\bvp)}/{\partial\lrs{n}}$ is given by
\begin{align}\nonumber
&\frac{\partial\Lambda(\bvp)}{\partial\lrs{n}}=
\sum_{k=1}^{\NL}\sum_{j\in\mtC} \frac{1}{\sigma_j^2} \int_{T^k_{1,j}}^{T^k_{2,j}}
\bigg(y^k_j(t)-\sum_{i\in\mtC}h^k_{j,i}\,s^k_i\big(t-\tau^k\big)\bigg)\\\label{eq:PartialSce1}
&\times\sum_{i\in\mtC}\bigg(\frac{{\partial h^k_{j,i}}}{{\partial \lrs{n}}} s^k_i(t - \tau^k) - h^k_{j,i} (s^k_i(t-\tau^k))'\frac{{\partial \tau^k}}{{\partial \lrs{n}}} \bigg) dt \end{align}
for $n\in\{1,2,3\}$. From \eqref{eq:PartialSce1}, the elements of the FIM in \eqref{eq:genericFIMsyc} can be obtained, after some manipulation, as follows:
\begin{align}\nonumber
	&[\Jv]_{n_1,n_2}=\sum_{k=1}^{\NL}\sum_{j\in\mtC} \frac{1}{\sigma_j^2}\int_{T^k_{1,j}}^{T^k_{2,j}}
\sum_{i\in\mtC}\sum_{l\in\mtC}\\\nonumber
&~\bigg(\frac{{\partial h^k_{j,i}}}{{\partial \lrs{n_1}}} s^k_i(t - \tau^k)
\frac{{\partial h^k_{j,l}}}{{\partial \lrs{n_2}}} s^k_l(t - \tau^k)
\\\nonumber&~-\frac{{\partial h^k_{j,i}}}{{\partial \lrs{n_1}}} s^k_i(t - \tau^k)
h^k_{j,l} (s^k_l(t-\tau^k))'\frac{{\partial \tau^k}}{{\partial \lrs{n_2}}}
\\\nonumber&~-\frac{{\partial h^k_{j,l}}}{{\partial \lrs{n_2}}} s^k_l(t - \tau^k)
h^k_{j,i} (s^k_i(t-\tau^k))'\frac{{\partial \tau^k}}{{\partial \lrs{n_1}}}
\\\label{eq:LastProp1}
&~+h^k_{j,i} (s^k_i(t-\tau^k))'\frac{{\partial \tau^k}}{{\partial \lrs{n_1}}}
h^k_{j,l} (s^k_l(t-\tau^k))'\frac{{\partial \tau^k}}{{\partial \lrs{n_2}}}\bigg)
\end{align}
As the signals $s_i^k(t-\tau^k)$ are assumed to be contained completely in the observation intervals $[T^k_{1,j},T^k_{2,j}]$, the expression in \eqref{eq:LastProp1} can be shown to be equal to that \eqref{eq:Jsyc1} based on the definitions in \eqref{eq:Eilpos}, \eqref{eq:Eil1pos}, and \eqref{eq:Eil2pos}. Also, the partial derivatives in \eqref{eq:tauDer} and \eqref{eq:alpDer} can be obtained from \eqref{eq:tau} with $\Delta^k=0$ for all $k\in\{1,\ldots,\NL\}$ and \eqref{eq:hji}, respectively. Overall, the CRLB on the MSE of any unbiased estimator $\lrh$ for the location of the VLC receiver, $\lr$, can be expressed via \eqref{crlb_expression} as $\rmE\big\{\|\lrh - \lr \|^2\big\}\geq {\rm{trace}}\big\{  \Jv^{-1} \big\}$. Since $\Jv$ in \eqref{eq:LastProp1} and $\Jsyn$ in \eqref{eq:Jsyc1} are equivalent, the
expression in \eqref{eq:CRLBsync} is reached.\hfill$\blacksquare$

Proposition~1 provides a closed-form expression for the CRLB on location estimation in VLP systems with RGB LEDs based on a generic three-dimensional setup. From the expression in \eqref{eq:Jsyc1}, it is noted that both the RSS information and the TOA information are utilized in Scenario~1, and their relative contributions depend on signal characteristics via the $E^{,k}_{i,l}$, $E'^{,k}_{i,l}$ and $E''^{,k}_{i,l}$ terms. Also, different signals (colors) emitted from each LED transmitter contribute to the localization accuracy, as expected. In the special case of a single color at each LED transmitter and a single PD at the VLC receiver, the FIM in Proposition~1 reduces to that in \cite[Prop.~1]{Direct_TCOM}.

\subsubsection{Scenario~2: Asynchronous System with Known Channel Attenuation Formula}\label{sec:Case2pos}

In this scenario, the VLC receiver is not synchronized with the LED transmitters, and the $\Delta^k$ terms in \eqref{eq:tau} are modeled as deterministic unknown parameters. Therefore, the TOA parameters are unknown and do not contribute to localization accuracy. Hence, the set of unknown parameters in \eqref{eqn:loglike} is specified as $\bvp=[\lrs{1}~\lrs{2}~\lrs{3}~\tau^1\cdots\,\tau^{\NL}]^T$ in this scenario. The CRLB in Scenario~2 is presented in the following proposition:

\textit{\textbf{Proposition~2}: In Scenario~2, the CRLB on the MSE of any unbiased estimator $\lrh$ for the location of the VLC receiver is expressed as}
\begin{gather}\label{eq:CRLBasyn}
\expectation \big\{ \big\|\lrh - \lr \big\|^2 \big\}\geq{\rm{trace}}\big\{\Jasyi\big\}
\end{gather}
\textit{where $\Jasy$ {denotes} a $3\times3$ matrix with the following elements:}
\begin{align}\nonumber
&[\Jasy]_{n_1,n_2}=\sum_{k=1}^{\NL}\Bigg(\sum_{j\in{\mtC}}\sum_{i\in{\mtC}}
\sum_{l\in{\mtC}}\frac{\partial h^k_{j,i}}{\partial \lrs{n_1}}\frac{\partial h^k_{j,l}}{\partial \lrs{n_2}}\frac{E^k_{i,l}}{\sigma_j^2}-
\\\nonumber&\bigg(\sum_{j\in{\mtC}}\sum_{i\in{\mtC}}\sum_{l\in{\mtC}}\frac{\partial h^k_{j,i}}{\partial \lrs{n_1}}\frac{h^k_{j,l}E'^{,k}_{i,l}}{\sigma_j^2}\bigg)
\bigg(\sum_{j\in{\mtC}}\sum_{i\in{\mtC}}\sum_{l\in{\mtC}}\frac{\partial h^k_{j,i}}{\partial \lrs{n_2}}\frac{h^k_{j,l}E'^{,k}_{i,l}}{\sigma_j^2}\bigg)
\\\label{eq:JasyExp}
&\Big{/}\bigg(\sum_{j\in{\mtC}}\sum_{i\in{\mtC}}\sum_{l\in{\mtC}}
\frac{h^k_{j,i}h^k_{j,l}E''^{,k}_{i,l}}{\sigma_j^2}\bigg)\Bigg)
\end{align}
\textit{for $n_1,n_2\in\{1,2,3\}$, with $E^k_{i,l}$, $E'^{,k}_{i,l}$, $E''^{,k}_{i,l}$, and ${\partial h^k_{j,i}}/{\partial \lrs{n}}$ being defined by \eqref{eq:Eilpos}, \eqref{eq:Eil1pos}, \eqref{eq:Eil2pos}, and \eqref{eq:alpDer}, respectively.}

\indent\indent\textit{Proof:} For the parameter vector given by $\bvp=[\lrs{1}~\lrs{2}~\lrs{3}~\tau^1\cdots\,\tau^{\NL}]^T$, the partial derivatives of the log-likelihood function in \eqref{eqn:loglike} are calculated as follows:
\begin{align}\nonumber
\frac{\partial\Lambda(\bvp)}{\partial\lrs{n}}&=\sum_{k=1}^{\NL}\sum_{j\in{\mtC}} \frac{1}{\sigma_j^2} \int_{T^k_{1,j}}^{T^k_{2,j}}
\bigg(y^k_j(t)-\sum_{i\in{\mtC}}h^k_{j,i}\,s^k_i\big(t-\tau^k\big)\bigg)\\\label{eq:PartialSce2}
&\times\sum_{i\in{\mtC}}\frac{{\partial h^k_{j,i}}}{{\partial \lrs{n}}} s^k_i(t - \tau^k) dt \\\nonumber
\frac{\partial\Lambda(\bvp)}{\partial\tau^k}&=-\sum_{j\in{\mtC}} \frac{1}{\sigma_j^2} \int_{T^k_{1,j}}^{T^k_{2,j}}
\bigg(y^k_j(t)-\sum_{i\in{\mtC}}h^k_{j,i}\,s^k_i\big(t-\tau^k\big)\bigg)\\\label{eq:PartialSce2_2}
&\times\sum_{i\in{\mtC}}h^k_{j,i} \big(s^k_i(t - \tau^k)\big)' dt
\end{align}
for $n\in\{1,2,3\}$ and $k\in\{1,\ldots,\NL\}$. From \eqref{eq:PartialSce2} and \eqref{eq:PartialSce2_2}, the FIM in \eqref{eq:fim} can be obtained as
\begin{gather}\label{eq:blockMat}
	\Jv=\begin{bmatrix}
		\JA&\JB\\\JBT&\JD
	\end{bmatrix}
\end{gather}
where $\JA$ is a $3\times3$ matrix with elements
\begin{align}\label{eq:FIM_RSS}
	[\JA]_{n_1,n_2} = \sum_{k=1}^{\NL}\sum_{j\in{\mtC}}\sum_{i\in{\mtC}}
\sum_{l\in{\mtC}}\frac{\partial h^k_{j,i}}{\partial \lrs{n_1}}\frac{\partial h^k_{j,l}}{\partial \lrs{n_2}}\frac{E^k_{i,l}}{\sigma_j^2}
\end{align}
for $n_1, n_2 \in \{1, 2, 3 \}$, $\JB$ is a $3\times\NL$ matrix with elements
\begin{align}\label{eq:FIM_RSS_2}
	[\JB]_{n,k} = -\sum_{j\in{\mtC}}\sum_{i\in{\mtC}}\sum_{l\in{\mtC}}\frac{\partial h^k_{j,i}}{\partial \lrs{n}}\frac{h^k_{j,l}E'^{,k}_{i,l}}{\sigma_j^2}
\end{align}
for $n \in \{1, 2, 3 \}$ and $k \in \{1, \ldots ,\NL\}$, and $\JD$ is an $\NL\times\NL$ matrix with elements
\begin{align}\label{eq:FIM_RSS_3}
	[\JD]_{k_1,k_2} = \begin{cases}
		\sum_{j\in{\mtC}}\sum_{i\in{\mtC}}\sum_{l\in{\mtC}}
\frac{h^k_{j,i}h^k_{j,l}E''^{,k}_{i,l}}{\sigma_j^2}\,, &{\textrm{if }}k_1=k_2\\
		0\,,&{\textrm{if }}k_1\ne k_2
	\end{cases}
\end{align}
for $k_1, k_2 \in \{1,\ldots ,\NL\}$. In \eqref{eq:FIM_RSS}--\eqref{eq:FIM_RSS_3}, $E^k_{i,l}$, $E'^{,k}_{i,l}$, $E''^{,k}_{i,l}$, and ${\partial h^k_{j,i}}/{\partial \lrs{n}}$ are as defined in \eqref{eq:Eilpos}, \eqref{eq:Eil1pos}, \eqref{eq:Eil2pos}, and \eqref{eq:alpDer}, respectively.

From \eqref{crlb_expression}, the CRLB on the location $\lr$ of the VLC receiver can be expressed as \cite{Poor}
\begin{gather}\label{eq:CRLBasynProof}
	\expectation \big\{ \big\|\lrh - \lr \big\|^2 \big\}\geq{\rm{trace}}\Big\{\big[\Jvi\big]_{3\times3}\Big\}
\end{gather}
where $\lrh$ is any unbiased estimator for $\lr$. From \eqref{eq:blockMat}, $\big[\Jvi\big]_{3\times3}$ can be computed as
\begin{gather}\label{eq:blockInv}
	\big[\Jvi\big]_{3\times3}=\left(\JA-\JB\JDi\JB^T\right)^{-1}.
\end{gather}
Since $\JD$ in \eqref{eq:FIM_RSS_3} is a diagonal matrix, the elements of $\JA-\JB\JDi\JB$ can be stated as
\begin{gather}\label{eq:ABDB}	\big[\JA-\JB\JDi\JB^T\big]_{n_1,n_2}=[\JA]_{n_1,n_2}-\sum_{k=1}^{\NL}\frac{[\JB]_{n_1,k}[\JB]_{n_2,k}}{[\JD]_{k,k}}
\end{gather}
By inserting \eqref{eq:FIM_RSS}--\eqref{eq:FIM_RSS_3} into \eqref{eq:ABDB}, the expression in \eqref{eq:JasyExp} is obtained. This observation together with \eqref{eq:CRLBasynProof} verifies the expressions in \eqref{eq:CRLBasyn} and \eqref{eq:JasyExp} in the proposition.\hfill$\blacksquare$

Proposition~2 presents a generic closed-form expression for the CRLB in Scenario~2, which illustrates that the location relation information in extracted only from the channel attenuation factors (RSS parameters) in this scenario as there exits no synchronization between the VLC receiver and the LED transmitters. The expression in Proposition~2 covers the CRLB expression in \cite[Prop.~3]{Direct_TCOM} as a special case when single-color LEDs and a VLC receiver with a single PD are employed.

\subsubsection{Scenario~3: Synchronous System with Unknown Channel Attenuation Formula}\label{sec:Case3pos}

In the last scenario, the LED transmitters and the VLC receiver are synchronized (i.e., $\Delta^k=0$ for all $k\in\{1,\ldots,\NL\}$ in \eqref{eq:tau}) but the VLC receiver does not know the channel attenuation formula in \eqref{eq:hji}. Therefore, only the TOA parameters contribute to localization accuracy, and the set of unknown parameters in \eqref{eqn:loglike} becomes $\bvp=\big[\lrs{1}~\lrs{2}~\lrs{3}~\{h^k_{j,i}\}_{k=1,j\in{\mtC},i\in{\mtC}}^{\NL}\big]^T$. Namely, there exist $9\NL+3$ unknown parameters. The CRLB in this scenario is provided in the following proposition:

\textit{\textbf{Proposition~3}: In Scenario~3, the CRLB on the MSE of any unbiased estimator $\lrh$ for the location of the VLC receiver can be stated as}
\begin{align}\label{eq:CRLBsce3}
&\expectation \big\{ \big\|\lrh - \lr \big\|^2 \big\}\geq{\rm{trace}}\big\{\JsyniT\big\}
\\\label{eq:Sce3blockFIM}
&\JsynT=\JAt-\JBt\JDti\JBt^T
\end{align}
\textit{where $\JAt$ is a $3\times3$ matrix with elements}
\begin{gather}\label{eq:JAtsce3}
\JAt=\left[\sum_{j\in{\mtC}}\frac{1}{\sigma_j^2}\sum_{k=1}^{\NL}\sum_{i\in{\mtC}}
\sum_{l\in{\mtC}}\frac{\partial \tau^k}{\partial \lrs{n_1}}\frac{\partial \tau^k}{\partial \lrs{n_2}}h^k_{j,l}h^k_{j,i}E''^{,k}_{i,l}\right]
\end{gather}
\textit{for $n_1,n_2\in\{1,2,3\}$, $\JBt$ is a $3\times9\NL$ matrix with elements}
\begin{gather}\label{eq:JBtsce3}
\JBt=\left[-\frac{1}{\sigma_j^2}\sum_{l\in{\mtC}}\frac{\partial \tau^k}{\partial \lrs{n}}h^k_{j,l}E'^{,k}_{i,l}\right]
\end{gather}
\textit{for $n\in\{1,2,3\}$ and $\{k,j,i\}\in\{1,\ldots,\NL\}\times\mtC\times\mtC$, and $\JDt$ is a $9\NL\times9\NL$ matrix with elements}
\begin{gather}\label{eq:JDtsce3}
\JDt=\left[\frac{E^k_{i,l}}{\sigma_j^2}\mathbbm{1}_{\{k=\tilde{k}, j=\tilde{j}\}}\right]
\end{gather}
\textit{for $\{k,j,i\}\in\{1,\ldots,\NL\}\times\mtC\times\mtC$ and $\{\tilde{k},\tilde{j},l\}\in\{1,\ldots,\NL\}\times\mtC\times\mtC$. In \eqref{eq:JAtsce3}--\eqref{eq:JDtsce3}, $E^k_{i,l}$, $E'^{,k}_{i,l}$, $E''^{,k}_{i,l}$, and ${\partial \tau^k}/{\partial \lrs{n}}$ are as defined by \eqref{eq:Eilpos}, \eqref{eq:Eil1pos}, \eqref{eq:Eil2pos}, and \eqref{eq:tauDer}, respectively, and $\mathbbm{1}_{\{k=\tilde{k}, j=\tilde{j}\}}$ denotes the indicator function.}

\indent\indent\textit{Proof:} In this scenario, the partial derivatives of the log-likelihood function in \eqref{eqn:loglike} are computed as
\begin{align}\nonumber
\frac{\partial\Lambda(\bvp)}{\partial\lrs{n}}&=-\sum_{k=1}^{\NL}\sum_{j\in{\mtC}} \frac{1}{\sigma_j^2} \int_{T^k_{1,j}}^{T^k_{2,j}}
\bigg(y^k_j(t)-\sum_{i\in{\mtC}}h^k_{j,i}\,s^k_i\big(t-\tau^k\big)\bigg)\\\label{eq:PartialSce3}
&\times\sum_{i\in{\mtC}}h^k_{j,i}\frac{{\partial \tau^k}}{{\partial \lrs{n}}} \big(s^k_i(t - \tau^k)\big)' dt \\
\frac{\partial\Lambda(\bvp)}{\partial h^k_{j,i}}&=\int_{T^k_{1,j}}^{T^k_{2,j}}
\bigg(y^k_j(t)-\sum_{i\in{\mtC}}h^k_{j,i}\,s^k_i\big(t-\tau^k\big)\bigg)
\label{eq:PartialSce3_2}
\frac{s^k_i(t - \tau^k)}{\sigma_j^2} dt
\end{align}
for $n\in\{1,2,3\}$ and $\{k,j,i\}\in\{1,\ldots,\NL\}\times\mtC\times\mtC$. After some manipulation, it can be derived from \eqref{eq:PartialSce3} and \eqref{eq:PartialSce3_2} that the FIM in \eqref{eq:fim} is in the form of
\begin{gather}\label{eq:blockMat_3}
	\Jv=\begin{bmatrix}
		\JAt&\JBt\\\JBt^T&\JDt
	\end{bmatrix}
\end{gather}
where $\JAt$, $\JBt$, and $\JDt$ are as in \eqref{eq:JAtsce3}, \eqref{eq:JBtsce3}, and \eqref{eq:JDtsce3}, respectively. Since $\expectation \big\{ \|\lrh - \lr \|^2 \big\}\geq{\rm{trace}}\big\{\big[\Jvi\big]_{3\times3}\big\}$ and $\big[\Jvi\big]_{3\times3}=\big(\JAt-\JBt\JDt^{-1}\JBt^T\big)^{-1}$, the expressions in the proposition are obtained.\hfill$\blacksquare$

Via Proposition~3, the theoretical accuracy limit on localization can be calculated for synchronized VLP systems with RGB LEDs, where the VLC receiver does not know the channel attenuation formula in \eqref{eq:hji} due to certain reasons such as unknown transmitter parameters or calibration problems. It should be noted that localization of VLC receivers in Scenario~3 has not been considered in the literature even in the special case of single-color LEDs and a VLC receiver with a single PD. In that special case, the CRLB can be calculated as in the following corollary.

\textit{\textbf{Corollary~1}: Suppose that each LED has a single color, say red, and the VLC receiver has a single PD for that color. Then, in Scenario~3, the CRLB on the MSE of any unbiased estimator $\lrh$ for the location of the VLC receiver is given by}
\begin{gather}\label{eq:CRLBsce3_Cor}
\expectation \big\{ \big\|\lrh - \lr \big\|^2 \big\}\geq{\rm{trace}}\big\{\JsyniT\big\}
\end{gather}
\textit{where $\JsynT$ is a $3\times3$ matrix with the following elements:}
\begin{align}\label{eq:Sce3blockFIMcor}
[\JsynT]_{n_1,n_2}=\sum_{k=1}^{\NL}\left(E''^{,k}_{r,r}
-\frac{\big(E'^{,k}_{r,r}\big)^2}{E^{k}_{r,r}}\right)\frac{\big(h^k_{r,r}\big)^2}{\sigma_r^2}
\frac{\partial \tau^k}{\partial \lrs{n_1}}\frac{\partial \tau^k}{\partial \lrs{n_2}}
\end{align}
\textit{for $n_1,n_2\in\{1,2,3\}$.}

\indent\indent\textit{Proof:} When each LED transmitter emits only the red color (signal) and the VLC receiver has a single PD for that color, the matrices $\JAt$, $\JBt$, and $\JDt$ in \eqref{eq:JAtsce3}, \eqref{eq:JBtsce3}, and \eqref{eq:JDtsce3} of Proposition~3 become
\begin{gather}\label{eq:JAtsce3cor}
\JAt=\left[\frac{1}{\sigma_r^2}\sum_{k=1}^{\NL}\frac{\partial \tau^k}{\partial \lrs{n_1}}\frac{\partial \tau^k}{\partial \lrs{n_2}}h^k_{r,r}h^k_{r,r}E''^{,k}_{r,r}\right]
\end{gather}
for $n_1,n_2\in\{1,2,3\}$,
\begin{gather}\label{eq:JBtsce3cor}
\JBt=\left[-\frac{1}{\sigma_j^2}\frac{\partial \tau^k}{\partial \lrs{n}}h^k_{r,r}E'^{,k}_{r,r}\right]
\end{gather}
for $n\in\{1,2,3\}$ and $k\in\{1,\ldots,\NL\}$, and
\begin{gather}\label{eq:JDtsce3cor}
\JDt=\left[\frac{E^k_{r,r}}{\sigma_r^2}\mathbbm{1}_{\{k=\tilde{k}\}}\right]
\end{gather}
for $k,\tilde{k}\in\{1,\ldots,\NL\}$. Then, the result in the corollary follows from the relations
$\expectation \big\{ \|\lrh - \lr \|^2 \big\}\geq{\rm{trace}}\big\{\big[\Jvi\big]_{3\times3}\big\}$ and $\big[\Jvi\big]_{3\times3}=\big(\JAt-\JBt\JDt^{-1}\JBt^T\big)^{-1}$ based on the expressions in \eqref{eq:JAtsce3cor}, \eqref{eq:JBtsce3cor}, and \eqref{eq:JDtsce3cor}.\hfill$\blacksquare$

\subsection{ML Estimators for Position Estimation}\label{sec:MLpos}

In this section, ML estimators are derived for localization of the VLC receiver in the scenarios considered above.

\subsubsection{ML Estimator for Scenario~1}\label{sec:ML1pos}

The ML estimator for the location of the VLC receiver in Scenario~1 is expressed as
\begin{gather}\label{eq:ML1pos}
\lrh=\underset{\lr}{\arg\max}~\Lambda(\bvp)
\end{gather}
where $\Lambda(\bvp)$ is given by \eqref{eqn:loglike} and $\bvp=\lr$. Based on the expression in \eqref{eqn:loglike}, the ML estimator in \eqref{eq:ML1pos} can be stated, after some manipulation, as follows:
\begin{align}\nonumber
\lrh=\underset{\lr}{\arg\max}~\sum_{k=1}^{\NL}\sum_{j\in{\mtC}}\frac{1}{\sigma_j^2}
&\bigg(
\sum_{i\in{\mtC}}h^k_{j,i}R_{y^k_j,s^k_i}\big(\tau^k\big)
\\\label{eq:ML1pos2}
&-0.5\sum_{i\in{\mtC}}\sum_{l\in{\mtC}}h^k_{j,i}h^k_{j,l}E^k_{i,l}
\bigg)
\end{align}
where $h^k_{j,i}$ and $\tau^k$ are functions of $\lr$ as in \eqref{eq:hji} and \eqref{eq:tau} (with $\Delta^k=0$ for all $k\in\{1,\ldots,\NL\}$), respectively, and
\begin{gather}\label{eq:RysPos}
R_{y^k_j,s^k_i}\big(\tau^k\big)\triangleq \int_{T^k_{1,j}}^{T^k_{2,j}}y^k_j(t)s^k_i\big(t-\tau^k\big)dt\,.
\end{gather}

It is noted from \eqref{eq:ML1pos2} that the ML estimator for Scenario~1 requires a three-dimensional search over all possible locations of the VLC receiver. For each possible location $\lr$, the correlation term $R_{y^k_j,s^k_i}\big(\tau^k\big)$ should be calculated for all $k\in\{1,\ldots,\NL\}$, $j\in\mtC$, and $i\in\mtC$ (i.e., $9\NL$ times), which constitutes the operation with the highest complexity. Hence, by considering the exhaustive search method (due to the non-convexity of the problem), the correlation terms should be calculated $9\NL\NV$ times in total, where $\NV$ denotes the number of possible values of $\lr$.
In addition, the ML estimator in \eqref{eq:ML1pos2} reduces to that in \cite[eqn.~(19)]{Direct_TCOM} in the special case of single-color LEDs and a VLC receiver with a single PD.

\subsubsection{ML Estimator for Scenario~2}\label{sec:ML2pos}

The ML estimator in Scenario~2 is given by
\begin{gather}\label{eq:ML2pos}
(\lrh,\btauh)=\underset{(\lr,\btau)}{\arg\max}~\Lambda(\bvp)
\end{gather}
where $\btau\triangleq\big[\tau^1\cdots\tau^{\NL}\big]^T$, $\bvp=(\lr,\btau)$, and $\Lambda(\bvp)$ is given by \eqref{eqn:loglike}. From \eqref{eqn:loglike}, \eqref{eq:ML2pos} can be expressed as
\begin{align}\nonumber
(\lrh,\btauh)=\underset{(\lr,\btau)}{\arg\max}~\sum_{k=1}^{\NL}\sum_{j\in{\mtC}}
&\frac{1}{\sigma_j^2}
\bigg(
\sum_{i\in{\mtC}}h^k_{j,i}R_{y^k_j,s^k_i}\big(\tau^k\big)
\\\label{eq:ML2pos2}
&-0.5\sum_{i\in{\mtC}}\sum_{l\in{\mtC}}h^k_{j,i}h^k_{j,l}E^k_{i,l}
\bigg)
\end{align}
where $h^k_{j,i}$ are functions of $\lr$ as in \eqref{eq:hji} and $R_{y^k_j,s^k_i}\big(\tau^k\big)$ is given by \eqref{eq:RysPos}. The ML estimator in \eqref{eq:ML2pos2} can also be implemented as follows:
\begin{align}\nonumber
\lrh=\underset{\lr}{\arg\max}~\sum_{k=1}^{\NL}\sum_{j\in{\mtC}}&\frac{1}{\sigma_j^2}
\bigg(
\sum_{i\in{\mtC}}h^k_{j,i}R_{y^k_j,s^k_i}\Big(\widehat{\tau}^{\,k}(\lr)\Big)
\\\label{eq:ML2pos3}
&-0.5\sum_{i\in{\mtC}}\sum_{l\in{\mtC}}h^k_{j,i}h^k_{j,l}E^k_{i,l}
\bigg)
\end{align}
where
\begin{gather}\label{eq:tauHat2}
\widehat{\tau}^{\,k}(\lr)=\underset{\tau^k}{\arg\max}~\sum_{j\in{\mtC}}\frac{1}{\sigma_j^2}
\sum_{i\in{\mtC}}h^k_{j,i}R_{y^k_j,s^k_i}\big(\tau^k\big)
\end{gather}
for $k\in\{1,\ldots,\NL\}$.

The ML estimator described by \eqref{eq:ML2pos3} and \eqref{eq:tauHat2} indicates that a three-dimensional search over all possible locations of the VLC receiver should be implemented together with $\NL$ one-dimensional searches for each possible location $\lr$. During each one-dimensional search in \eqref{eq:tauHat2}, the correlation term $R_{y^k_j,s^k_i}\big(\tau^k\big)$ should be calculated for all possible delay values $\tau^k$ (considering exhaustive search). If $N_{\tau^k}$ denotes the number of possible values for $\tau^k$, the correlation terms should be calculated $9\NL\sum_{k=1}^{\NL}N_{\tau^k}$ times for each $\lr$, and $9\NL\NV\sum_{k=1}^{\NL}N_{\tau^k}$ times in total (with $\NV$ denoting the number of possible values of $\lr$). Hence, the ML estimator in Scenario~2 has higher complexity than that in Scenario~1 (see \eqref{eq:ML1pos2}).

In the special case of single-color LEDs and a VLC receiver with a single PD \cite[Sec.~IV-B]{Direct_TCOM}, the one-dimensional search in \eqref{eq:tauHat2} becomes independent of the VLC location $\lr$. In that case, the complexity of the ML estimation in \eqref{eq:ML2pos3} and \eqref{eq:tauHat2} reduces significantly. Namely, a three-dimensional search over $\lr$ is performed and, in total, the correlation terms are calculated $\sum_{k=1}^{\NL}N^k_{\tau}$ times only.

\subsubsection{ML Estimator for Scenario~3}\label{sec:ML3pos}

The ML estimator in Scenario~3 is formulated as
\begin{gather}\label{eq:ML3pos}
\Big(\lrh,\big\{\widehat{h}^k_{j,i}\big\}_{k=1,j\in{\mtC},i\in{\mtC}}^{\NL}\Big)=
\underset{\big(\lr,\{h^k_{j,i}\}_{k=1,j\in{\mtC},i\in{\mtC}}^{\NL}\big)}{\arg\max}~\Lambda(\bvp)
\end{gather}
where $\bvp=\big[\lrs{1}~\lrs{2}~\lrs{3}~\{{h}^k_{j,i}\}_{k=1,j\in{\mtC},i\in{\mtC}}^{\NL}\big]^T$ and $\Lambda(\bvp)$ is as in \eqref{eqn:loglike}. From \eqref{eqn:loglike}, \eqref{eq:ML3pos} can be expressed as (cf.~\eqref{eq:ML3_2} and \eqref{eq:ML1pos2})
\begin{align}\label{eq:ML3pos2}
\Big(\lrh,\big\{\hjkh\big\}_{k=1,j\in{\mtC}}^{\NL}\Big)
&=\underset{\big(\lr,\{\hjk\}_{k=1,j\in{\mtC}}^{\NL}\big)}
{\arg\max}\sum_{k=1}^{\NL}\sum_{j\in{\mtC}}\frac{1}{\sigma_j^2}
\\\nonumber
&\times\bigg(\big(\hjk\big)^T\Rjk\big(\tau^k\big)-0.5\big(\hjk\big)^T\Ek\hjk\bigg)
\end{align}
where $\tau^k$ are functions of $\lr$ as in \eqref{eq:tau} (with $\Delta^k=0$ for all $k\in\{1,\ldots,\NL\}$),
\begin{align}\label{eq:ML3posDefs}
\hjk&\triangleq[h^k_{j,r}~h^k_{j,g}~h^k_{j,b}]^T\\
\Rjk\big(\tau^k\big)&\triangleq\big[R_{y^k_j,s^k_r}\big(\tau^k\big)~R_{y^k_j,s^k_g}\big(\tau^k\big)
~R_{y^k_j,s^k_b}\big(\tau^k\big)\big]^T\\
\Ek&\triangleq\begin{bmatrix}
E^k_{r,r}&E^k_{r,g}&E^k_{r,b}\\
E^k_{g,r}&E^k_{g,g}&E^k_{g,b}\\
E^k_{b,r}&E^k_{b,g}&E^k_{b,b}
\end{bmatrix}
\end{align}
The objective function in \eqref{eq:ML3pos2} is a quadratic expression in terms of $\hjk$, and $\Ek$ is positive semi-definite by definition. Hence, the gradients with respect to $\hjk$ can be set to zero to characterize the ML estimator as follows:
\begin{gather}\label{eq:gradZero}
\nabla_{\hjk}\Lambda(\bvp)=\frac{1}{\sigma_j^2}
\big(\Rjk\big(\tau^k\big)-\Ek\hjk\big)=\bf{0}
\end{gather}
for $k\in\{1,\ldots,\NL\}$ and $j\in{\mtC}$. Assuming that $\Ek$ is invertible, the relation in \eqref{eq:gradZero} becomes $\hjk=\big(\Ek\big)^{-1}\Rjk\big({\tau}^k\big)$, which reduces the ML estimator in \eqref{eq:ML3pos2} to the following problem (cf.~\eqref{eq:ML3_3}):
\begin{align}\label{eq:ML3pos3}
\lrh
=\underset{\lr}
{\arg\max}\sum_{k=1}^{\NL}\sum_{j\in{\mtC}}\frac{1}{2\sigma_j^2}
\big(\Rjk\big({\tau}^k\big)\big)^T
\big(\Ek\big)^{-1}
\Rjk\big({\tau}^k\big)
\end{align}

The ML estimator in \eqref{eq:ML3pos3} for Scenario~3 requires a three-dimensional search over all possible locations of the VLC receiver, and $9\NL$ evaluations of the correlation terms $R_{y^k_j,s^k_i}\big(\tau^k\big)$ for each $\lr$. Hence, the complexity order of the ML estimator in Scenario~3 is the same as that in Scenario~1.


\section{Numerical Results}\label{sec:Nume}

\subsection{Simulation Setup}\label{sec:NumeSetup}

\begin{figure}
        \centering
        \includegraphics[width=.98\linewidth]{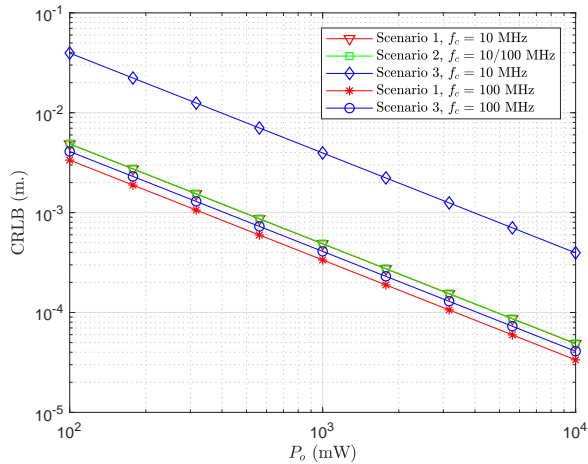}
        \caption{CRLB vs. $P_o$ for $x = 5$\,m. and $T_s = 0.01\,$sec.}
\label{fig:CRLBvsPo}
\end{figure}

\begin{figure}
        \centering
        \includegraphics[width=.98\linewidth]{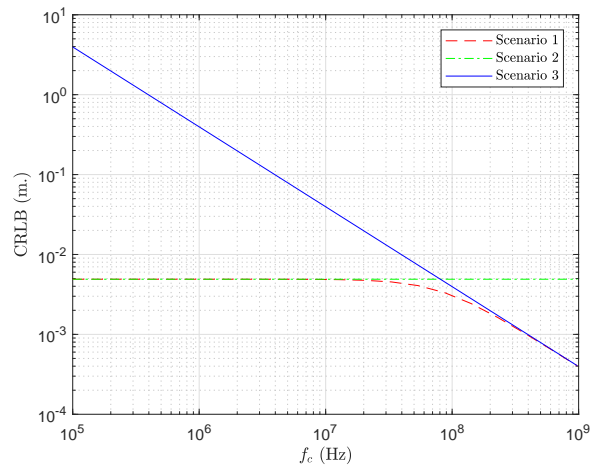}
        \caption{CRLB vs. $f_c$ for $x=5$\,m., $T_s=0.01\,$sec., $P_o=0.1$\,W}
        \label{fig:CRLBvsFc}
\end{figure}

\begin{figure}
        \centering
        \includegraphics[width=.98\linewidth]{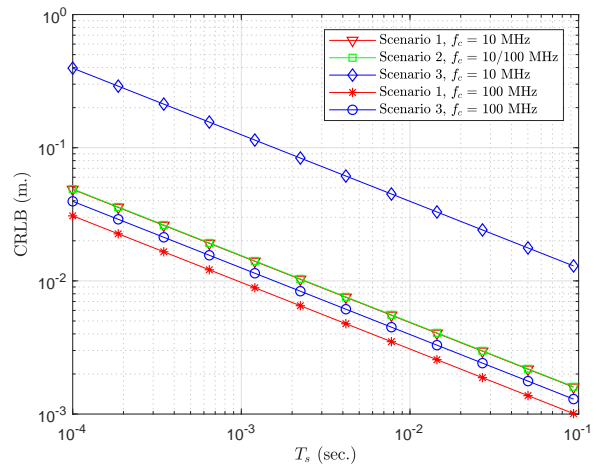}
        \caption{CRLB vs. $T_s$ for $x = 5$\,m. and $P_o=0.1$\,W.}
        \label{fig:CRLBvsTs}
\end{figure}

\begin{figure}
        \centering
        \includegraphics[width=.98\linewidth]{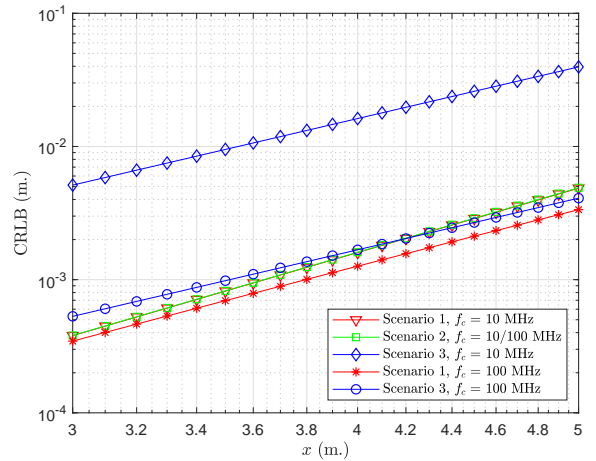}
        \caption{CRLB vs. $x$ for $P_o=0.1$\,W and $T_s = 0.01\,$sec.}
        \label{fig:CRLBvsX}
\end{figure}


In this section, numerical examples are presented to investigate the CRLBs derived in Sections~\ref{sec:Limits}~and~\ref{sec:LimitsPos} and the performance of the ML estimators in Sections~\ref{sec:ML}~and~\ref{sec:MLpos}. For distance estimation, a similar setting to that in \cite{MFK_CRLB} is considered; that is, the Lambertian order is set to $m=1$ and $\tilde{h}$ in \eqref{eq:hji_2} is taken as $2.5$ meters. The areas of the PDs at the VLC receiver are set to $A_j=1\,$cm$^2$ for $j\in\mtC$, and the spectral density level of the noise components at different branches of the VLC receiver are $\sigma_j^2=1.336\times10^{-22}\,$W/Hz for $j\in\mtC$ \cite{MFK_CRLB,CRB_TOA_VLC}. For position estimation, a similar setting to that in \cite{Direct_TCOM} is analyzed. We consider a room with width, depth, and height of $[8~8~5]$~m., respectively, where $N_L=4$ LED transmitters are attached to the ceiling at positions $\lt{1}=[2~2~5]^T$, $\lt{2}=[6~2~5]^T$, $\lt{3}=[2~6~5]^T$, and $\lt{4}=[6~6~5]^T$ m. The orientation vectors of the LEDs in \eqref{eq:hji} are expressed as
\begin{gather}\label{eq:orient_res}
\nt{k} = [\sin\theta_k \cos\phi_k~\sin\theta_k \sin\phi_k~\cos\theta_k]^T
\end{gather}
for $k = 1,\dots,N_L$, where $\theta_k$ and $\phi_k$ denote the polar and azimuth angles, respectively \cite{Lampe_MISO_TSP_2016}. We consider the following angle configuration for the transmitters: $(\theta_1, \phi_1)=(150^{\circ},45^{\circ})$, $(\theta_2, \phi_2)=(150^{\circ},135^{\circ})$, $(\theta_3, \phi_3)=(150^{\circ},-45^{\circ})$, $(\theta_4, \phi_4)=(150^{\circ},-135^{\circ})$. The VLC receiver is located at $\lr=[4~4~1]^T$~m. looking upwards, i.e., its orientation vector is given by $\nr~=~[0~0~1]^T$  \cite{Direct_TCOM}. The transmitted signals from the LEDs are modeled as \cite{CRB_TOA_VLC}:
\begin{gather}\label{eq:st}
s_i^k(t)=P_o\left(1-\cos\left(\frac{2\pi t}{T_s}\right)\right)\left(1+\cos(2\pi f_i^k t)\right)
\end{gather}
for $t\in[0,T_s]$, $k=1,\dots,N_L$, and $i\in\mtC$, where $f_i^k$ is the center frequency for the $i$th signal (color) coming from transmitter $k$. $f_i^k$'s are specified through a constant center frequency $f_c$ as $f_i^k = k f_i$, where $f_r=0.9f_c$, $f_g=f_c$, and $f_b=1.1f_c$. Note that in the distance estimation problem, we consider the scenario where there is only one transmitter, i.e., $N_L=1$, and drop the index $k$ in the relevant definitions (implicitly setting $k=1$). Parameter $P_o$ in \eqref{eq:st} corresponds to the average emitted optical power (i.e., source optical power). In addition, the $\tilde{R}_{j,i}$ terms in \eqref{eq:hji} are taken as $[\tilde{R}_{r,r}~\tilde{R}_{r,g}~\tilde{R}_{r,b}]=0.4\times[1~0.042~0.03]$, $[\tilde{R}_{g,r}~\tilde{R}_{g,g}~\tilde{R}_{g,b}]=0.4\times[0.194~0.665~0.277]$, and $[\tilde{R}_{b,r}~\tilde{R}_{b,g}~\tilde{R}_{b,b}]=0.4\times[0.009~0.084~0.421]$, where $0.4\,$mA/mW represents a coefficient related to the responsivity of the PDs as in \cite{MFK_CRLB}, and the remaining numbers are adopted from eqn.~(14) in \cite{ColorShiftKeying_JLT14}.

\begin{figure}
\centering
\includegraphics[width=.98\linewidth]{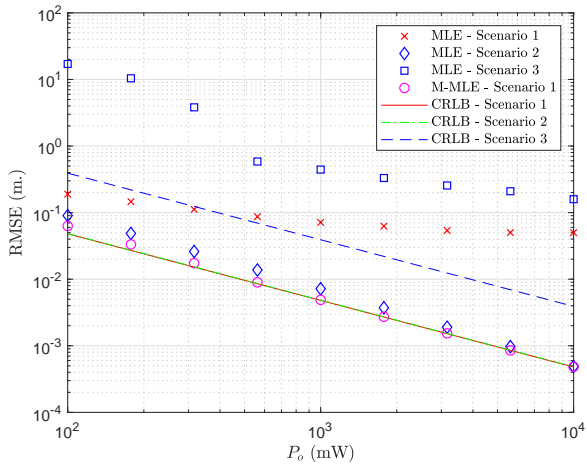}
\caption{RMSEs of ML estimators (MLEs) for distance estimation in different scenarios, together with the CRLBs, where $x=5\,$m., $f_c=10\,$MHz, and $T_s=0.1\,$ms., where M-MLE denotes the modified ML estimator in Section~\ref{sec:ML4}.}
\label{fig:RMSEs}
\end{figure}

\subsection{Distance Estimation}\label{sec:DistanceRes}

\begin{figure}
\centering
\includegraphics[width=.98\linewidth]{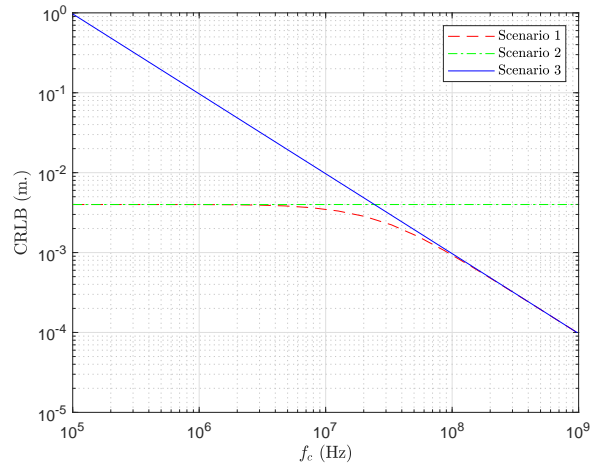}
\caption{CRLB vs. $f_c$ for $T_s = 0.01\,$sec. and $P_o = 0.1$\,W for position estimation.}
\label{fig:CRLBvFCpos}
\end{figure}

First, the CRLBs (in meters) for the considered scenarios in Section~\ref{sec:Limits} are plotted in Fig.~\ref{fig:CRLBvsPo} with respect to $P_o$ in \eqref{eq:st} (equivalently, with respect to source optical power), where $x = 5$\,m. and $T_s = 0.01\,$sec. It is noted that for small center frequencies (around $10\,$MHz), the CRLBs in Scenario~1 and Scenario~2 are almost the same since synchronization does not bring any additional benefits in this case. In other words, the distance related information contained in the RSS parameter is more significant than that in the TOA parameter. This can also be verified from the high CRLB values in Scenario~3 for $f_c=10\,$MHz as only the TOA parameter is utilized in that scenario. As the center frequencies are increased, the TOA parameter becomes significant and the CRLB in Scenario~3 decreases rapidly. Since only the RSS information is used in Scenario~2, its CRLB does not depend on the center frequencies. On the other hand, the CRLB of Scenario~1 also decreases with increased center frequencies as it utilizes both the RSS and TOA parameters in distance estimation.

\begin{figure}
\centering
\includegraphics[width=.98\linewidth]{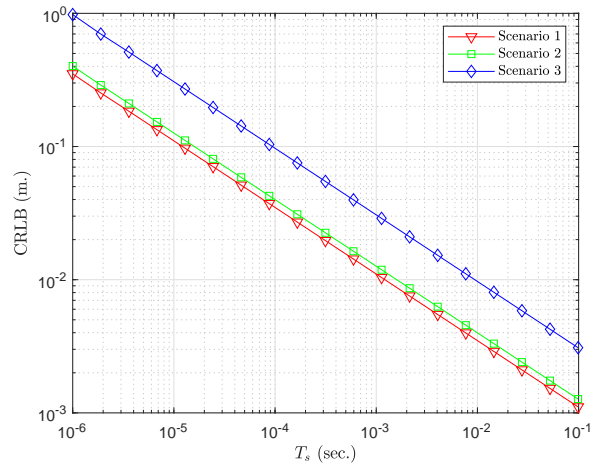}
\caption{CRLB vs. $T_s$ for $P_o = 0.1$\,W and $f_c = 10\,$MHz for position estimation.}
\label{fig:CRLBvTSpos}
\end{figure}

Fig.~\ref{fig:CRLBvsFc} illustrates the frequency dependencies of the CRLB expressions more explicitly, where $x = 5$\,m., $T_s = 0.01\,$sec., and $P_o=0.1$ are used. As the center frequencies of the transmitted signals are raised, the distance related information in the TOA parameter increases. Hence, the CRLB in Scenario~3, which only utilizes the TOA parameter, decreases with the center frequency parameter $f_c$ in Fig.~\ref{fig:CRLBvsFc}. On the other hand, the CRLB in Scenario~2 does not change with the center frequencies, as noted before. Since both the TOA and RSS parameters are utilized in Scenario~1, the CRLB is almost constant for small $f_c$'s (as the distance related information in the TOA parameter is insignificant compared to that in the RSS parameter in that region) and then starts decreasing with $f_c$ (as the distance related information in the TOA parameter gets significant).

Next, Fig.~\ref{fig:CRLBvsTs} presents the CRLB versus $T_s$ curves in the considered scenarios for two different center frequencies, where $x = 5$\,m. and $P_o=0.1$. As expected, the CRLB decreases as the duration $T_s$ of the transmitted signals in \eqref{eq:st} increases. In addition, the relative CRLB performances in different scenarios carry similarities to those in Fig.~\ref{fig:CRLBvsPo} due to the same reasons.

Moreover, the CRLBs are plotted versus the distance $x$ in Fig.~\ref{fig:CRLBvsX}, where $P_o=0.1$ and $T_s = 0.01\,$sec. As channel attenuation becomes more severe as the distance increases (see \eqref{eq:hji}), the CRLBs increase with distance. As expected, it is observed that the CRLBs increase with distance. However, the slopes of the CRLBs with respect to distance are not the same. The slope of the CRLB in Scenario~2 is higher than that in Scenario~3 since they are proportional to $x^{m+4}$ and $x^{m+3}$, respectively (considering the CRLBs in meters) based on the expressions in Sections~\ref{sec:Case2} and \ref{sec:Case3}. On the other hand, the slope of the CRLB in Scenario~1 (see \eqref{eq:CRLB1_3}) is almost the same as that in Scenario~2 for low center frequencies (as the RSS parameter is dominant in that case) and it is close to and higher than that in Scenario~3 for high center frequencies (as the TOA parameter is significant in that case, as well).

Furthermore, the root mean-squared errors (RMSEs) of the ML estimators derived in Section~\ref{sec:ML} are plotted versus $P_0$, together with the CRLBs, where $x=5\,$m., $f_c=10\,$MHz, and $T_s = 0.1\,$ms. From Fig.~\ref{fig:RMSEs}, it is observed that the RMSEs of the ML estimators in Scenario~1 (see \eqref{eq:ML1_2}) and Scenario~3 (see \eqref{eq:ML3_3}) are significantly higher than the corresponding CRLBs. The main reason for this is the finite sampling interval used in the simulations (namely, $0.5\,$ns), which limits the utilization of  distance related information contained in the TOA parameter (please see \cite{MFK_CRLB} for a similar observation). On the other hand, the ML estimator in Scenario~2 (see \eqref{eq:tauHat} and \eqref{eq:x2Hat}) and the modified ML estimator in Scenario~1 (see \eqref{eq:x4Hat}) achieve close performance to the CRLBs. The best performance is achieved in Scenario~1 as both the TOA and RSS parameters are utilized.

\subsection{Position Estimation}\label{sec:PositionRes}

Position estimation is performed in a room with the setup described in Section~\ref{sec:NumeSetup} by considering the scenarios specified in Section~\ref{sec:PosEst}. Figs.~\ref{fig:CRLBvFCpos} and \ref{fig:CRLBvTSpos} present the CRLBs for the position estimation problem with respect to the center frequency parameter $f_c$ (for $T_s=0.01\,$sec.) and the observation interval $T_s$ (for $f_c=10\,$MHz), respectively, where $P_o=0.1\,$W. We make similar observations to those for the distance estimation simulations in Section~\ref{sec:DistanceRes}. Namely, for lower values of the center frequency parameter, the CRLBs in Scenario 1 and Scenario 2 are almost identical since the information contained in the TOA parameter is inconsiderable compared to the RSS parameter, and the CRLB for Scenario 3 is very high. As the center frequency increases, Scenario 1 and Scenario 3 induce lower CRLBs since they exploit the information contained in the TOA parameter whereas Scenario 2 has a constant CRLB since the information in the TOA parameter is not utilized. Also, the CRLBs in all the scenarios decrease as the observation interval of the signals, $T_s$, increases.

\begin{figure}
\centering
\includegraphics[width=.98\linewidth]{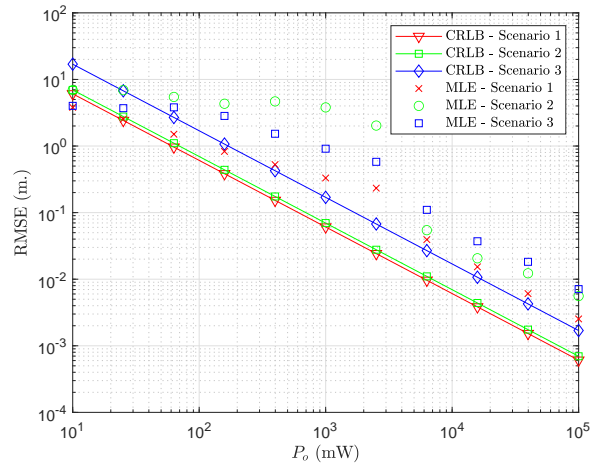}
\caption{RMSEs of ML estimators for position estimation, together with the CRLBs, where $f_c=10\,$MHz and $T_s=1 \mu\,$s.}
\label{fig:MLEvCRLBpos}
\end{figure}

Finally, we obtain the RMSEs of the ML estimators derived in Section~\ref{sec:MLpos} and present them together with the CRLBs in Section~\ref{sec:LimitsPos} in Fig.~\ref{fig:MLEvCRLBpos}, where $f_c=10\,$MHz and $T_s=1 \mu\,$s. Since $f_c$ is not very high, the CRLB in Scenario~3, where only the TOA information is utilized, is the highest for all source optical powers in compliance with the previous results. In addition, at high source optical powers, the ML estimators achieve RMSEs close to the CRLBs and the RMSEs are ordered in the same way as the CRLBs. On the other hand, for low and medium source optical powers, the CRLBs do not provide tight bounds on the RMSEs of the ML estimators (as expected) and the highest RMSEs are obtained in Scenario~2. Moreover, it is noted that the RMSEs can be lower than the CRLBs for low source optical powers since the search for the position of the VLC receiver is performed in the specified room whereas the CRLB derivations do no assume any prior information about the position of the VLC receiver.


\section{Concluding Remarks}\label{sec:Conc}

Performance limits and ML estimators have been derived for distance and position estimation in VLP systems in the presence of RGB LEDs by considering three different scenarios. In Scenario~1 and Scenario~2, a synchronous and an asynchronous system have been assumed, respectively, with a known channel attenuation formula at the VLC receiver. In Scenario~3, synchronism has been assumed but the channel attenuation formula has been modeled as unknown. Since both the TOA and RSS parameters are utilized in Scenario~1, it has the lowest CRLBs in all the cases. On the other hand, Scenario~2 achieves lower (higher) CRLBs than Scenario~3 for low (high) center frequencies (more generally, effective bandwidths).

The results obtained for distance estimation in Section~\ref{sec:DistEst} generalize the CRLBs and ML estimators in \cite{MFK_CRLB} to VLP systems with RGB LEDs and corresponding PDs. In addition, the CRLBs and ML estimators were derived in \cite{Direct_TCOM} for a single (white) LED at each transmitter and a single PD at the VLC receiver by considering Scenario~1 and Scenario~2. The results in Section~\ref{sec:PosEst} have not only extended the results in \cite{Direct_TCOM} to VLP systems with RGB LEDs but also covered a new scenario (Scenario~3) that has not been investigated for VLP systems before in the literature.



\bibliographystyle{IEEEtran}

\bibliography{Ref_proc_IEEE}

\end{document}